\documentclass{article}
\usepackage[utf8]{inputenc}

\usepackage[text={6.8in,8in},centering]{geometry}
\usepackage{amsmath}
\usepackage{amssymb}
\usepackage{amsthm}
\usepackage[english]{babel}
\usepackage{graphicx}
\usepackage{subfig} 
\usepackage{tikz}
\usepackage{url}
\usepackage{ulem}
\usepackage{bbold}
\usepackage{hyperref}
\usepackage{todonotes}
\usepackage{authblk}
\usepackage{color}

\newtheorem{lemma}{Lemma}[section]
\newtheorem{theorem}[lemma]{Theorem}

\newtheorem{corollary}[lemma]{Corollary }

\title{Tight Lieb-Robinson Bound for approximation ratio in Quantum Annealing}

\author[1,2]{Arthur Braida}
\author[2]{Simon Martiel}
\author[1]{Ioan Todinca}
\affil[1]{LIFO - Laboratoire d'Informatique Fondamentale d'Orléans, Université d'Orléans, France}
\affil[2]{Eviden Quantum Lab, Les Clayes-sous-Bois, France}

\date{February 2024}

\begin{document}

\maketitle

\begin{abstract}
    Quantum annealing (QA) holds promise for optimization problems in quantum computing, especially for combinatorial optimization. This analog framework attracts attention for its potential to address complex problems. Its gate-based homologous, QAOA with proven performance, has brought lots of attention to the NISQ era. Several numerical benchmarks try to compare these two metaheuristics however, classical computational power highly limits the performance insights. In this work, we introduce a parametrized version of QA enabling a precise 1-local analysis of the algorithm. We develop a tight Lieb-Robinson bound for regular graphs, achieving the best-known numerical value to analyze QA locally. Studying MaxCut over cubic graph as a benchmark optimization problem, we show that a linear-schedule QA with a 1-local analysis achieves an approximation ratio over 0.7020, outperforming any known 1-local algorithms. 
\end{abstract}


\section{Introduction}

Quantum annealing (QA), firstly introduced by \cite{Apolloni_1989,Kadowaki_1998}, is one of the most promising quantum algorithms to solve optimization problems \cite{PhysRevA.108.042403,Yarkoni_2022}. It runs on the quantum analog computational framework. Known as adiabatic quantum computing (AQC), it was  coined by Fahri et. al \cite{farhi2000quantum} in 2000 and stands for the analog part of the gate-based model. 
Although the two frameworks are known to be equivalent (one can efficiently simulate the other) \cite{albash2018adiabatic}, their studies rely on different theoretical tools. QA has gained lots of attention in the last decade because it seems well-suited to solve combinatorial optimization problems. One largely studied gate-based algorithm, namely QAOA \cite{farhi2014quantum}, is QA-inspired and has brought a lot of attention to the NISQ era \cite{Farhi2022quantumapproximate, Herrman2021MultiangleQA, shaydulin2019}. The goal of quantum annealing is to let a quantum system evolve along a trajectory according to the Schrödinger equation subject to a problem-dependent Hamiltonian. \\

\begin{figure}[ht]
    \centering
    \includegraphics[scale=0.34]{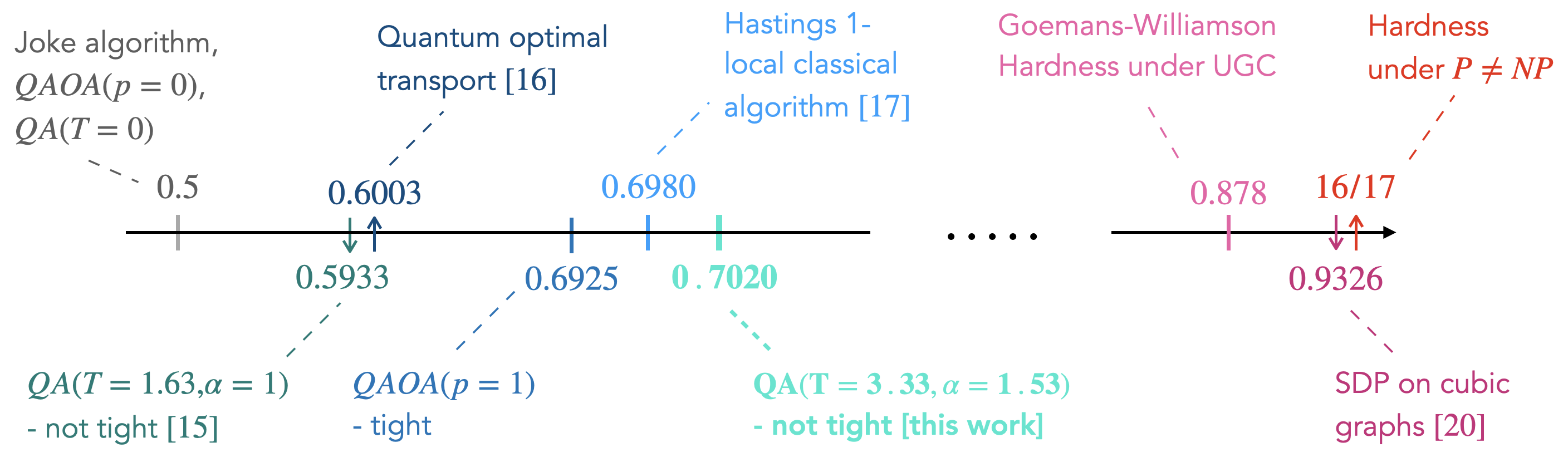}
    \caption{Some approximation ratios for MaxCut over cubic graphs. Only blue/greenish colored values result from 1-local algorithms. The highest proved ratio for MaxCut over cubic graphs is presented in \cite{Halperin2002MAXCI}.}
    \label{fig:approxcubic}
\end{figure}

A recent study from \cite{lykov2023sampling} suggests that QAOA, even in the NISQ era may bring a quantum advantage over classical algorithms for approximation in optimization problems. Several numerical studies like \cite{Pelofske_2023} suggest that QA performs well compared to QAOA. However, numerical studies in QA are rarely convincing because the size of the instance is limited by the classical computational power required to solve the Schrödinger equation or by the largest available quantum annealer.  The downside of the approach is that, due to the limited size of the input data for the numerical experiments, it is not possible to deduce a reliable asymptotic scale. To tackle this comparison, some researchers tried to develop new mathematical tools to derive an analytical bound on the algorithmic performance of QA. As it has been widely used to benchmark metaheuristics, we choose to focus on the approximation ratio of MaxCut over cubic (3-regular) graphs \cite{blekos2023review}. With one layer, standard QAOA achieves a ratio of 0.6925 \cite{farhi2014quantum}, i.e., for any input, the algorithm outputs a solution whose number of cut edges is at least 0.6925 times the number of edges cut by the optimal solution. A zoo of variants \cite{blekos2023review} of this algorithm has been proposed since but as in the original QAOA, we use the same initial and final Hamiltonians. A recent study in \cite{benchasattabuse2023lower} gives lower bounds on the number of rounds needed for some variants of QAOA to achieve a given ratio. In \cite{braida2022constant}, the authors manage to formally prove that constant time QA achieves a ratio of 0.5933 and, based on empirical studies, conjectured that the actual ratio is 0.6963. This last result has been improved by Banks, Brown and Warburton in \cite{banks2023rapid} to 0.6003. In \cite{hastings2019classical}, Hastings shows a simple classical local algorithm that outperforms QAOA with an approximation ratio of 0.6980 (Fig. \ref{fig:approxcubic}). By local algorithm, we mean that the decision of the algorithm on each node only depends on a ball of nodes at a constant distance from that node. 
The difficulty in comparing QA to QAOA or other local algorithms is that QA is non-local by nature. However, we can still make a local analysis of it. In 1972, Lieb and Robinson \cite{lieb1972finite} showed the existence of an information speed limit inside a quantum system. 
This means that, for a short enough evolution time $T$, the correlation between two remote sites of the quantum system decreases exponentially fast with respect to their distance $q$, typically in $\mathcal{O}(\frac{T^q}{q!})$.
This observation can be used to analyze QA as a local algorithm by taking into account this exponential decrease in correlation strength. We refer the reader to \cite{chen2023speed} for a recent extensive review on the Lieb-Robinson bound. 

\begin{figure}[ht]
    \centering
    \includegraphics[scale=0.35]{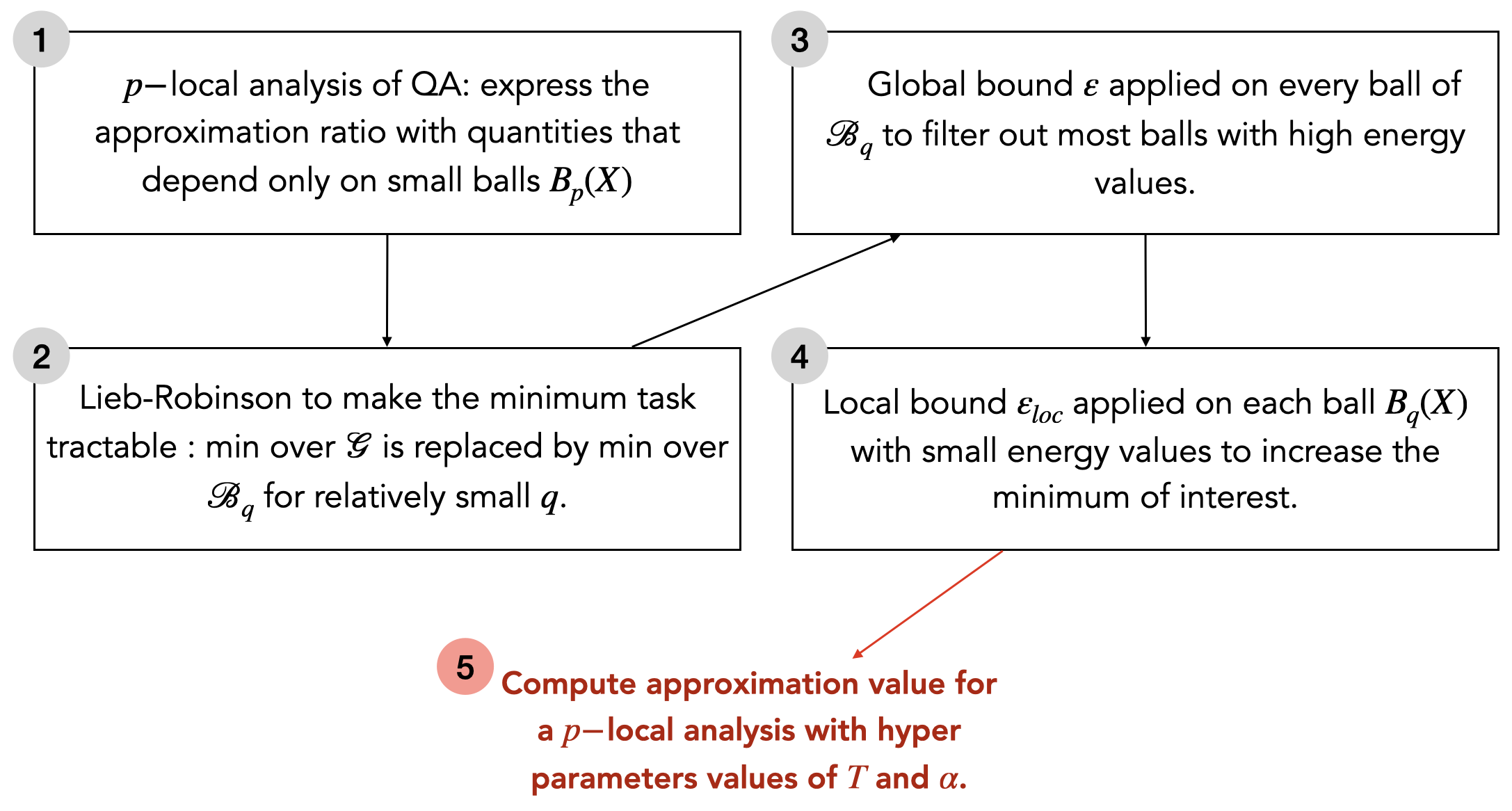}
    \caption{Overview of the analysis steps.}
    \label{fig:overview}
\end{figure}

In this work, we develop a super tight Lieb-Robinson (LR) bound by using the commutativity graph construction from \cite{Wang_2020} over general regular graphs. This LR bound is adapted to linear-schedule QA applied to MaxCut as we will define in the next Section \ref{sec:2}. In that setting, previously known state-of-the-art LR bounds \cite{Haah_2021, chen2023speed} achieve interesting numerical values for $q \simeq 100$ when ours only needs $q=3$ to reach similar values.
We also slightly modify the standard initial Hamiltonian with a free parameter $\alpha$ in front of it. 
It appears that this additional degree of freedom in the algorithm (formulation) allows for a tighter performance analysis.
Eventually, we end up proving that a 1-local analysis of QA brings the approximation ratio above 0.7020. This value shows that constant time QA outperforms both single-layer QAOA and the best-known classical 1-local algorithm. Schedule optimization should bring further improvements, and we have suggested one with a cubic function. The global overview of the different steps used to derive an approximation ratio is detailed in Fig. \ref{fig:overview}. 

The first step consists of deriving a formula of the ratio that depends only on local ball $B_p(X)$ centered on some edge $X$ of a regular graph $G$. By local ball, we mean the subgraph of $G$ generated by adding edges and nodes around $X$ up to distance $p$ from $X$ (see Section \ref{sec:2} for formal definition).
In step 2, we turn a worst-case analysis over all possible graphs $\mathcal{G}$ into a worst-case analysis over a finite set of relatively small graphs $\mathcal{B}_q$ thanks to a Lieb-Robinson bound.
Steps 3 and 4 deal with finding the minimum of interest by filtering the different balls after applying the global bound $\varepsilon$ and the local one $\varepsilon_{loc}$ on the leftovers. Eventually, these steps allow us to compute a numerical value for the approximation ratio depending on the running time $T$ and the free parameter $\alpha$. 

\paragraph{Organization of the paper:}In Section~\ref{sec:2}, we formally define the notion of approximation ratio, the parametrized family of $QA$ we will study, the notion of $p-$local analysis of it, and present how a Lieb-Robinson bound can be used to compute the numerical value of the ratio. In Section~\ref{sec:3}, we state the main theorem of the paper and show the derivation of the global and local LR bounds. In particular in Subsection~\ref{ssec:app}, we perform steps 3,4 and 5 of Fig. \ref{fig:overview} to finish the proof of the theorem. We hint a possible improvement with a non-linear schedule. In Section~\ref{sec:discussion}, we discuss this result and the meaning of the parameter $\alpha$, introduced for the tightness of the analysis. Although the construction is problem-dependent, we give some insight on the generalization of the construction for other degree, schedule and problems.

\section{Local analysis of QA and Lieb-Robinson Bound}
\label{sec:2}
In this section, we define the approximation ratio which attests to the performance of QA and allows us to compare it with other algorithms. We then introduce the parametrized version of the QA process we will analyze and define a $p-$local analysis in the case of quantum annealing. Eventually, by introducing the Lieb-Robinson bound, we explain how QA can be locally analyzed in order to compare it to QAOA. 

\paragraph{Approximation ratio:} Given an input graph $G$ on which we want to solve, an optimization problem $C$, and a probabilistic algorithm $\mathcal{A}$ that outputs a solution $x$ for the problem $C$ applied to $G$. One way to qualify the performance is to look at the average value reached by the output distribution of algorithm $\mathcal{A}$, written $\mathbb{E}_{\mathcal{A}}[C(G)]$. The metric that is used is the ratio of the latter quantity normalized by the optimal cost value $C_{opt}(G)$ for the specific input. In practice, we are interested in the worst case scenario and define the \textit{approximation ratio} as: 
\begin{equation}
    \rho_{C,\mathcal{A}}=\min_G\frac{\mathbb{E}_{\mathcal{A}}[C(G)]}{C_{opt}(G)}
\end{equation} 

For the rest of this work, we will focus on quantum annealing to solve a particular optimization problem called MaxCut over cubic graphs. The goal of MaxCut problem is to find a bi-partition that maximizes the number of edges across the bi-partition.. Unless stated otherwise, we fix $\mathcal{A}=QA$ and we will remove its dependency on the variables notation for clarity. MaxCut is a well-known problem in fundamental computer science and has several applications in physics and electrical circuits \cite{Barahona1988}. It admits a natural encoding into a Hamiltonian that has the same interaction topology as the graph instance.

\paragraph{Parametrized QA:} Given a graph $G(V, E)$ in $\mathcal{G}$, a family of graph, on which we want to solve MaxCut, the target hamiltonian is $H_1^G=-\sum_{X\in E}O_X$ where $O_X=\frac{1-\sigma_z^{(a)}\sigma_z^{(b)}}{2}$ for the edge $X=(a,b)$.   With a minus, this Hamiltonian encodes $-C$, so the goal is to minimize this function. Starting from the uniform superposition $|\psi_0\rangle$ we are interested in the mean of $H_1^G$ of the final state $|\psi^G(T,\alpha)\rangle=U_{T,\alpha}^G|\psi_0\rangle$. Here, $U_{T,\alpha}^G$ denotes the unitary evolution operator under the time-dependent Hamiltonian $H(t,G)=(1-\frac{t}{T})H_0(\alpha)+\frac{t}{T}H_1^G$ and $T$ the total annealing time. We chose a parameterized initial Hamiltonian $H_0(\alpha)=-\sum_i \frac{\sigma_x^{(i)}}{\alpha}$, where $\alpha>0$. This operator $U_{T,\alpha}^G$, corresponds to a Schrödinger evolution, i.e. is a solution of:
$$
\forall t \in [0,T], \quad i\hbar \frac{dU_{t,\alpha}^G}{dt}=H(t,G)U_{t,\alpha}^G
$$

The expected value at the end of the annealing process is $\langle H_1^G\rangle_{G,\alpha}=\langle \psi^G(T,\alpha)| H_1^G|\psi^G(T,\alpha)\rangle$ and thus, our metric of interest is 
$$\rho_{MC} = \max_{T,\alpha} \min_{G \in \mathcal{G}} \frac{-\langle H_1^G\rangle_{G,\alpha}}{C_{opt}(G)}.$$ 

Since we are interested in using local arguments to bound this quantity, we will restrict the family of graphs $\mathcal{G}$ to the set of 3-regular graphs. The goal is to find a good lower bound for this ratio. This can be achieved by separately upper bounding the denominator and lower bounding the numerator. By linearity, the numerator can be written as a sum over the edges $\sum_{X \in E}\langle O_X \rangle_{G,\alpha}$.

\paragraph{$p-$local analysis:} Due to its analog nature, QA is non local, so to compare it to other optimization solver like QAOA or to local classical algorithms, we need to define a way to analyze it as a local algorithm. For any $X \in E(G)$ and any positive integer $p$, we note $B_p(X)$ the subgraph composed by nodes and edges situated on paths of length at most $p$ from any endpoint of $X$. We note $\mathcal{B}_p$ the set of all possible balls $B_p(X)$ over all graphs $G \in \mathcal{G}$ and all possible edges $X$. We call a $p-$local analysis of QA, an analysis that produces an approximation ratio that depends only on balls in $\mathcal{B}_p$. Let us develop the construction of a 0-local and 1-local analysis. \\

\noindent \textit{0-local: }For any $X \in E$, a trivial lower bound of the summands in the numerator of the ratio is $\min_{G \in \mathcal{G}} \langle O_X \rangle_G=\langle O_X \rangle_\mathcal{G}$, where $\mathcal{G}$ is the set of all 3-regular graphs. With the trivial bound on $C_{opt} < |E|$, it gives the following lower bound on the approximation ratio: $$\rho_{MC}\geq \max_{T,\alpha} \langle O_X \rangle_\mathcal{G}.$$ 
Finding the latter value constitutes the approximation ratio of QA with 0-local analysis.  \\ 

\begin{figure}[ht]
    \centering
    \includegraphics[scale=0.4]{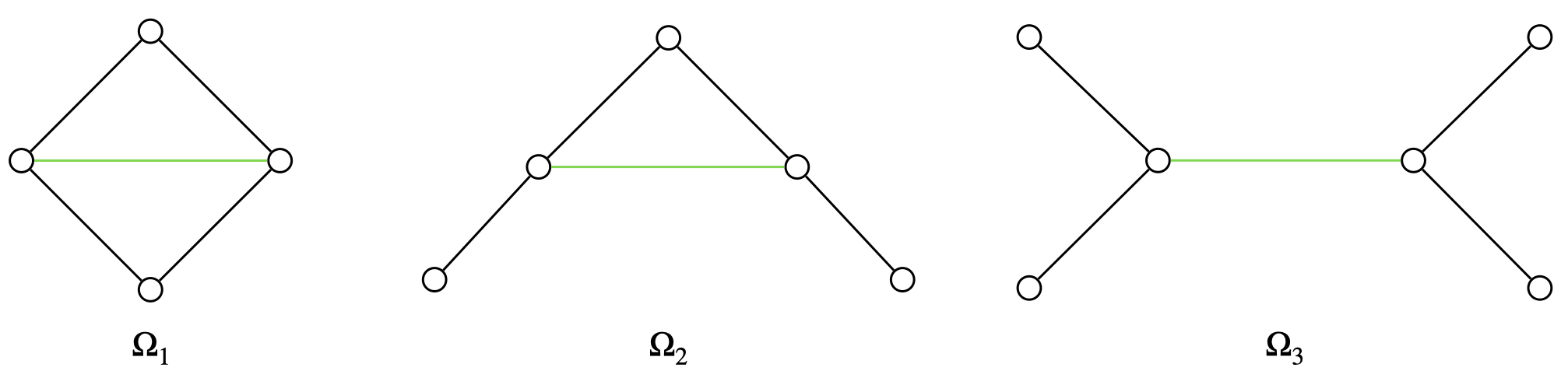}
    \caption{All possible balls $B_1(X)$ for $X$ an edge (in green) in a 3-regular graph.}
    \label{fig:omegas}
\end{figure}

\noindent \textit{1-local: }To improve this bound, we will consider the neighboring structure of each edge. For a $p=1$-local analysis, we look at $B_1(X)$, the ball of radius 1 around $X$. In 3-regular graphs, there are three such $B_1(X)$, $\Omega_1$ the square, $\Omega_2$ the triangle and $\Omega_3$ the double binary tree (see Fig. \ref{fig:omegas}).

Coming back to our edge energies $\langle O_X \rangle_G$ for all $X \in E$: we can bound these terms by one of the following quantities $ \langle O_X \rangle_{\mathcal{G}}^{\Omega_i} = \min \{\langle O_X \rangle_G \mid G \in \mathcal{G}, X \in E(G) \text{ s.t. } B_1(X) = \Omega_i\} $. 
Let us define $n_i$ as the number of configurations $\Omega_i$ in $G$. Using the regularity constraints, we have that the number $n_3$ of edges in configuration $\Omega_3$ is $n_3 = |E|- 5n_1 - 3n_2$ (see \cite{farhi2014quantum,braida2022constant} for more details). The final expected energy can thus be lower bounded as:
$$
-\langle H_1^G\rangle_{G} \geq n_1 \langle O_X \rangle_{\mathcal{G}}^{\Omega_1}+(4n_1+3n_2) \langle O_X \rangle_{\mathcal{G}}^{\Omega_2}+(|E|-5n_1-3n_2) \langle O_X \rangle_{\mathcal{G}}^{\Omega_3} 
$$ This expression still depends on the input graph through the variables $n_i$, and thus, still needs to be minimized over the positive integers $n_i$'s with the constraint that $4n_1+3n_2\leq |V|$. The upper bound on the optimal value $C_{opt}$ can be refined by observing that at least one edge is uncut in the MaxCut solution in configuration $\Omega_1$ and $\Omega_2$, i.e. $C_{opt} \leq |E|-n_1-n_2$, where $|E|=3|V|/2$. This gives the following lower bound for the approximation ratio:
\begin{equation}
\label{eq:min1loc}
    \rho_{MC}\geq \max_{T,\alpha} \min_{n1,n2 \text{ st } 4n_1+3n_2 \leq |V|} \frac{n_1 \langle O_X \rangle_{\mathcal{G}}^{\Omega_1}+(4n_1+3n_2) \langle O_X \rangle_{\mathcal{G}}^{\Omega_2}+(|E|-5n_1-3n_2) \langle O_X \rangle_{\mathcal{G}}^{\Omega_3} }{|E|-n_1-n_2}
\end{equation} 
Given a quite loose condition on the value of $\langle O_X \rangle_{\mathcal{G}}^{\Omega_i}$ which is satisfied in our case (see Appendix \ref{app:min} and \ref{ssec:app}), the minimization gives $n_1=n_2=0$ and the approximation ratio becomes $\rho_{MC}\geq \max_{T,\alpha} \langle O_X \rangle_{\mathcal{G}}^{\Omega_3}$. However, as in the 0-local analysis, the difficulty lies in computing the minimum over all graphs to find the values of $\langle O_X \rangle_{\mathcal{G}}^{\Omega_i}$. Even for 3-regular graphs, because $\mathcal{G}$ is infinite, this minimum is intractable. \\

\paragraph{Lieb-Robinson bound:} QA is \textit{a priori} non-local, however Lieb and Robinson demonstrated that the information flow has a bounded speed inside a quantum system. In other words, limiting the amount of time during which the quantum system evolves also limits the distance at which two sites can strongly correlate. Using this on a QA process allows us to analyze QA locally. For a graph $G(V,E)$, the idea is to approximate $\langle O_X \rangle_G$ by the energy of $\langle O_X \rangle_{B_q(X)}$. This step will ease the minimization task as now $\mathcal{B}_q$ is a finite family of graphs.\\

\noindent \textit{Local bound: }Suppose that for any graph $G$, and any edge $X$ of $G$, there exists a $\varepsilon_{loc}(B_q(X),T,\alpha)>0$ that upper bounds the absolute difference: $|\langle O_X \rangle_G - \langle O_X \rangle_{B_q(X)}|<\varepsilon_{loc}(B_q(X),T,\alpha)$. The local aspect lies in the fact that $\varepsilon_{loc}$ depends on the ball $B_q(X)$. If such a bound exists, we have that $\langle O_X \rangle_G \geq \langle O_X \rangle_{B_q(X)} - \varepsilon_{loc}(B_q(X),T,\alpha)$. This lower bound is satisfied for all graphs $G \in \mathcal{G}$. The only dependence on the input graph now lies in $B_q(X)$. We can rewrite the minimization over $\mathcal{G}$ as:
\begin{align}
    &\min_{G \in \mathcal{G}} \langle O_X \rangle_G \geq \min_{B_q(X) \in \mathcal{B}_q} \left (\langle O_X \rangle_{B_q(X)} - \varepsilon_{loc}(B_q(X),T,\alpha) \right ) \\
    \label{eq:LRmin}
    \Rightarrow & \langle O_X \rangle_{\mathcal{G}} \geq  \langle O_X \rangle_{\mathcal{B}_q}^*
\end{align} where $\langle O_X \rangle_{\mathcal{B}_q}^*=\min_{B_q(X) \in \mathcal{B}_q} \left (\langle O_X \rangle_{B_q(X)} - \varepsilon_{loc}(B_q(X),T,\alpha) \right )$. Therefore, the approximation ratio becomes,  for QA as a 0-local algorithm,
\begin{equation}
    \rho_{MC}\geq \max_{T,\alpha} \langle O_X \rangle_{\mathcal{B}_q}^*
\end{equation}
We can do the same for the 1-local analysis, when taking advantage of the neighborhood at radius 1 around the edge $X$. So we have that 
\begin{align}
\label{eq:minBq}
    \langle O_X \rangle_{\mathcal{G}}^{\Omega_i} \geq  \min_{B_{q}}(X) \in \mathcal{B}_q \left (\langle O_X \rangle_{B_q(X)}^{\Omega_i} - \varepsilon_{loc}(B_q(X),T,\alpha)\right )= \langle O_X \rangle_{\mathcal{B}_{q,i}}^*
\end{align}
where $\mathcal{B}_{q,i}$ is the family of graphs $B_q(X) \in \mathcal{B}_{q}$ restricted to balls $B_q(X)$ for which $B_1(X)=\Omega_i$.
The approximation ratio can be written as a new equation that depends only on these epsilons and worst edge energy among “small” ball of radius $q$:
\begin{equation}
\label{eq:minrho}
    \rho_{MC}\geq \max_{T,\alpha} \min_{n1,n2 \text{ st } 4n_1+3n_2 \leq |V|} \frac{n_1 \langle O_X \rangle_{\mathcal{B}_{q,1}}^*+(4n_1+3n_2) \langle O_X \rangle_{\mathcal{B}_{q,2}}^*+(|E|-5n_1-3n_2) \langle O_X \rangle_{\mathcal{B}_{q,3}}^* }{|E|-n_1-n_2}
\end{equation}
This method only helps the minimization task if $q$ is small enough. Indeed, the size of $\mathcal{B}_q$ grows exponentially with $q$ and and so does the size of ball. Computing $\langle O_X \rangle_{\mathcal{B}_q}$ requires to solve the Schrodinger equation on balls of size up to $\mathcal{O}(2^q)$. For cubic graphs, $q=3$ seems reasonable: all balls in $\mathcal{B}_3$ plus all the regular ones in $\mathcal{B}_2$ amounts for 930449 balls.
However, we still need to have a large enough $p$ so that the $\varepsilon_{loc}(B_q(X),T,\alpha)$ quantities are small enough to not degrade our lower bound \ref{eq:minrho} (typically, we want $\varepsilon_{loc}(B_q(X),T,\alpha)\approx 10^{-2}$). Using state-of-the-art generic LR bounds \cite{Haah_2021,chen2023speed} would require considering balls of radius $q\approx 100$ in order to reach such a precision. Section \ref{sec:3} is dedicated to the derivation of a tighter LR bound tailored exactly for the purpose of having reasonable $\varepsilon_{loc}(B_q(X),T,\alpha)$ values for $q=3$. \\

\noindent \textit{Global bound: }To avoid having to develop too many bounds $\varepsilon_{loc}$ as the number of $B_q(X)$ explodes exponentially, it can be useful to apply a global bound for most of the balls in the minimization task. We define it as
\begin{equation}
    \varepsilon(q,T,\alpha)= \max_{B_q(X) \in \mathcal{B}_q} \varepsilon_{loc}(B_q(X),T,\alpha)
\end{equation}
We will see in Section \ref{sec:3} that this maximum is easy to derive from the analytical expression of the local bound. The global bound is used as follows:
\begin{align*}
    \left (\langle O_X \rangle_{B_q(X)} - \varepsilon_{loc}(B_q(X),T,\alpha) \right ) 
    &\geq  \langle O_X \rangle_{B_q(X)} - \max_{B_p(X) \in \mathcal{B}_q}\varepsilon_{loc}(B_q(X),T,\alpha)  \\
    &= \langle O_X \rangle_{B_q(X)} - \varepsilon(q,T,\alpha) \\
    \Rightarrow \langle O_X \rangle_{\mathcal{B}_q}^* &\geq \langle O_X \rangle_{\mathcal{B}_q} - \varepsilon(q,T,\alpha)
\end{align*}
Then for balls $B_q(X)$ for which $\langle O_X \rangle_{B_q(X)}$ is large enough, the global bound is sufficient in the minimization task over $\mathcal{B}_q$. \\

\noindent In this sections we introduced different concepts and their role for obtaining the approximation ratio of QA solving MaxCut on cubic graphs, based on local analysis. In the next section we formally prove that the approximation ratio of QA for MaxCut on cubic graphs is greater than $0.7020$. For that purpose we provide tighter LR bounds than the existing ones in the literature, leading to explicit numerical values for quantities $\varepsilon$, $\varepsilon_{loc}$ and $\langle O_X \rangle^*_{\mathcal{B}_q}$.

\section{Tight LR bound on regular graphs for MaxCut approximation}
\label{sec:3}

The statement of our main result is the following:
\begin{theorem}
\label{thm:main}
    With a 1-local analysis, the approximation ratio reached by $QA$ to solve MaxCut over cubic graphs is above 0.7020.
\end{theorem}

In this section we prove Theorem \ref{thm:main} in two steps. First we develop a tight enough LR bound using the commutativity graph structure introduced in \cite{Wang_2020} and by computing the exact values of the schedule's nested integrals. This determines the required value of $q$ to achieve the best provable ratio. The second step is {purely numerical and requires} to enumerate each ball $B_q(X)$ and get the minimum of the final expected energy of the edge $X$ inside these balls for each $\Omega_i$, corrected by the LR bound. Our approach follows the algorithm presented in the last section of \cite{braida2022constant}.

\subsection{Tight LR bound on regular graphs in QA}
As mentioned in section \ref{sec:2}, the minimization to obtain the approximation ratio is intractable when performed over the entire graph family. However, the LR bound helps to reduce the size of the set on which we have to minimize to a finite subfamily of graphs, namely $\mathcal{B}_q$ (see Equation \ref{eq:LRmin}). \\

\noindent First we seek to develop a local bound $\varepsilon_{loc}(B_q(X),T,\alpha)$ such that $|\langle O_X \rangle_G - \langle O_X \rangle_{B_q(X)}|<\varepsilon_{loc}(B_q(X),T,\alpha)$ for all $d$-regular graphs $G$ such that the ball at distance $q$ around edge $X$ corresponds to $B_q(X)$. Let $G$ be such a graph. It happens to be very difficult to manipulate this expression considering the initial state $|\psi_0\rangle $ so the first step (however costly in terms of tightness) is to get rid of this dependency by working directly with the evolution operator: $$ |\langle O_X \rangle_G - \langle O_X \rangle_{B_q(X)}| \leq \left\lVert O_X^G(T) - O_X^{B_q(X)}(T)\right\rVert$$ where we introduce the evolved observable under $U_T^G$ and $U_T^{B_q(X)}$ respectively, once again dropping the $\alpha$ in the notation for clarity. They are defined as:
\begin{align}
    O_X^G(T) &= (U_T^G)^\dagger O_X U_T^G \\
    O_X^{B_q(X)}(T) &= (U_T^{B_q(X)})^\dagger O_X U_T^{B_q(X)} 
\end{align} 
In \cite{Bravyi_2006}, the authors demonstrate that the evolution over a subset of nodes can also be expressed as :

\begin{align*}
    O_X^{B_q(X)}(T) 
    &= \int d \mu (U) U O_X^G(T) U^\dagger
\end{align*} 
where $\mu(U)$ denotes the Haar measure over unitary operator $U$ supported on $S=V \backslash V(B_p(X))$. Noticing that $ U O_X^G(T) U^\dagger = O_X^G(T) + U[O_X^G(T), U^\dagger]$, we can bound the quantity of interest $\left\lVert O_X^G(T) - O_X^{B_q(X)}(T) \right\rVert \leq \int d \mu (U) \left\lVert [O_X^G(T), U] \right\rVert $ for any unitary $U$ supported on $S$. We are then left to bound the norm of this commutator. 

Let us introduce a helpful tool presented in \cite{Wang_2020}: the \textit{commutativity graph} $\mathbf{G}(\mathbf{V},\mathbf{E})$ associated to a Hamiltonian $H(t,G)$ having local interactions. In general, we can write $H(t) = \sum_j h_j(t) \gamma_j$ where $\gamma_j$ are hermitian operators with norm less than unity and $h_j(t)$ are time-dependent scalars. The commutativity graph of $H(t)$ is constructed such that each operator $\gamma_j$ is represented by a node $j$ and two nodes $j$ and $k$ are connected if and only if $[\gamma_j,\gamma_k] \neq 0$. The structure of the graph captures the commutative and non-commutative relationships between the operators in the Hamiltonian.

In the case of MaxCut, the total time-dependent Hamiltonian writes 
$$
H(t, G) = \sum_{i \in V} (1-\frac{t}{T})\frac{\sigma_x^{(i)}}{\alpha}+\sum_{(a,b)\in E}\frac{t}{T}\frac{1-\sigma_z^{(a)}\sigma_z^{(b)}}{2}
$$ 
As described, the terms $\gamma_j$ of the Hamiltonian are represented as nodes in the commutativity graph $\mathbf{G}$. We can distinguish two types of nodes in $\mathbf{V}$: those corresponding to interaction operators over the edges $E$ of the original input graph, and those corresponding to local operators over nodes of $V$. This means that we have for $e=(a,b)\in E$, $\gamma_e=\frac{1-\sigma_z^{(a)}\sigma_z^{(b)}}{2}$ and for $v\in V$, $\gamma_v=\frac{\sigma_x^{(v)}}{\alpha}$. We can rewrite the total Hamiltonian as:

$$H(t, G) = \sum_{v \in V}  h_{v}(t)\gamma_{v} +  \sum_{e\in E} h_{e}(t)\gamma_{e} $$

\begin{figure}[ht]
    \centering
    \includegraphics[scale=0.4]{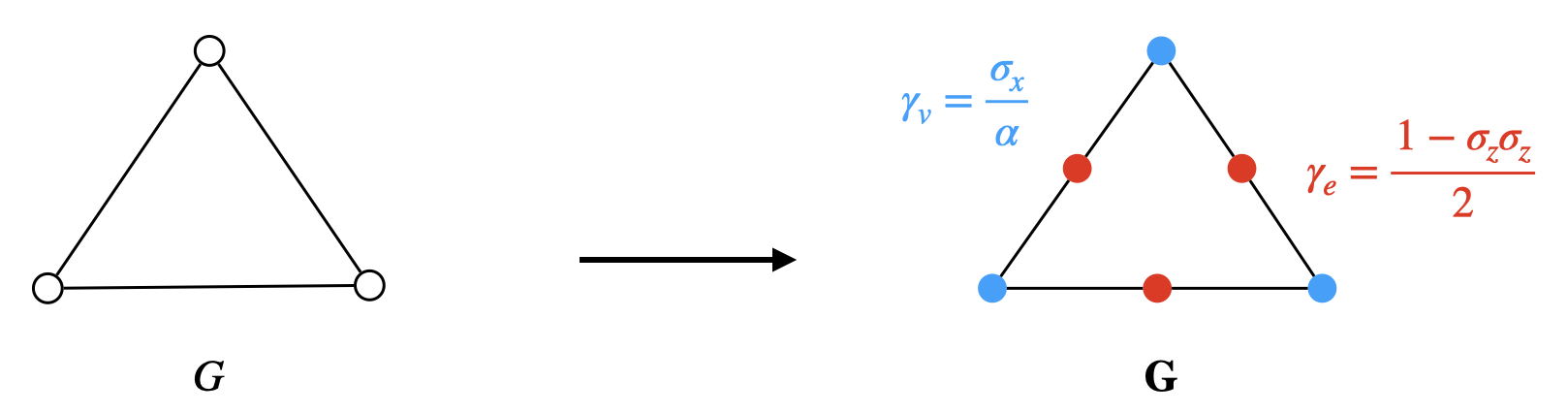}
    \caption{Example of a commutativity graph for the following Hamilonian: $H(t,G)= \sum_{i=1}^3 (1-\frac{t}{T}) \frac{\sigma_x^{(i)}}{\alpha} + \frac{t}{T} \frac{1-\sigma_z^{(i)}\sigma_z^{(i+1)}}{2}$ where index $i$ is taken modulo 3. Blue nodes represent 1-local operators and red nodes represent 2-interaction terms. In particular, blue nodes of $\mathbf{G}$ correspond to nodes of the original graph $G$, and red nodes correspond to its edges.}
    \label{fig:exComG}
\end{figure}

Our notation fixes the time-dependent scalars at $h_{e}(t)=\frac{t}{T}$ for $e\in E$ and $h_{v}(t)=1-\frac{t}{T}$ for $v \in V$. Also, it is obvious to see that the commutativity graph is bipartite. The only pairs that do not commute are pairs $(\gamma_v,\gamma_e)_{v,e}$, where node $v$ is incident to edge $e$ in $G$. An example of commutativity graph is shown in Fig. \ref{fig:exComG}.

For a unitary $A$ supported on $S$, we want to upper bound the quantity $ \left\lVert [O_X^G(T), A] \right\rVert$. This edge $X$ can be identified to a specific interaction term in the commutativity graph. Let us define $X$ to be the node in $\mathbf{G}$ corresponding to the edge $X$ and consider the operator $\gamma_{X}^A(T)=[\gamma_{X}(T),A]$ with $\gamma_{X}(t)=(U_T^G)^\dagger \gamma_{X}U_T^G$, dropping the dependency on $G$. Still following the first steps in \cite{Wang_2020}, we can arrive at a similar expression in the time-dependent regime (see Appendix \ref{app:proof} for details):

\begin{align}
\label{eq:normUP}
    \lVert \gamma_{X}^A(T) \rVert - \lVert \gamma_{X}^A(0) \rVert &\leq \sum_{v:\langle Xv \rangle \in \mathbf{G}}\int_0^T h_{v}(t) \left \lVert \left [ \gamma_{X}(t),\gamma_{v}^A(t) \right ] \right \rVert dt \nonumber \\
    &\leq \sum_{v:\langle Xv \rangle \in \mathbf{G}} \int_0^T (1-\frac{t}{T}) \left \lVert \gamma_{v}^A(t) \right \rVert dt
\end{align}

Now, we see on the right hand side of Eq. (\ref{eq:normUP}) that we have the norm of $\gamma_v(t)$ for some node $v$ adjacent to $X$ in $\mathbf{G}$. We can derive two update rules which we will use alternately depending on the considered node of $\mathbf{G}$, i.e. depending whether it corresponds to an edge $e$ of $G$ or to a node $v$ of $G$. These two rules are as follows: 

\begin{align}
    \lVert \gamma_{{e}}^A(t) \rVert - \lVert \gamma_{{e}}^A(0) \rVert &\leq \sum_{v:\langle e v \rangle \in \mathbf{G}}\int_0^t (1-\frac{t'}{T}) \left \lVert \gamma_{v}^A(t') \right \rVert dt' \\
    \label{eq:update_maxcut}
    \lVert \gamma_{v}^A(t) \rVert - \lVert \gamma_{v}^A(0) \rVert &\leq \frac{2}{\alpha}\sum_{e:\langle v e \rangle \in \mathbf{G}}\int_0^t \frac{t'}{T} \left \lVert \gamma_{e}^A(t') \right \rVert dt'
\end{align} where we used two inequalities that for any $t$ and any $U$: 
\begin{align*}
   \lVert \gamma_{e}^U(t) \rVert &=  \lVert [\gamma_{e}(t) , U]\rVert \leq \frac{1}{2}2\lVert \sigma_z^{(i)} \sigma_z^{(j)} \rVert \lVert U \rVert \leq \lVert U \rVert \\
   \lVert \gamma_{v}^U(t) \rVert &=  \lVert [\gamma_{v}(t) , U]\rVert \leq \frac{1}{\alpha}2\lVert \sigma_x \rVert \lVert U \rVert \leq \frac{2}{\alpha}  \lVert U \rVert
\end{align*} where $(i,j)$ is the edge on which $\gamma_{e}$ is applied and we used the trivial commutation of the identity with any matrix. Note that $\lVert \gamma_{j}^A(0) \rVert = 0$ as long as $j$ is inside $B_{q=k-1}(X)$, so we can iterate up to $2k$ steps as the first node outside $B_q(X)$ is a red node corresponding to an interaction term (see Fig. \ref{fig:comGraph}):

\begin{figure}[ht]
    \centering
    \includegraphics[scale=0.5]{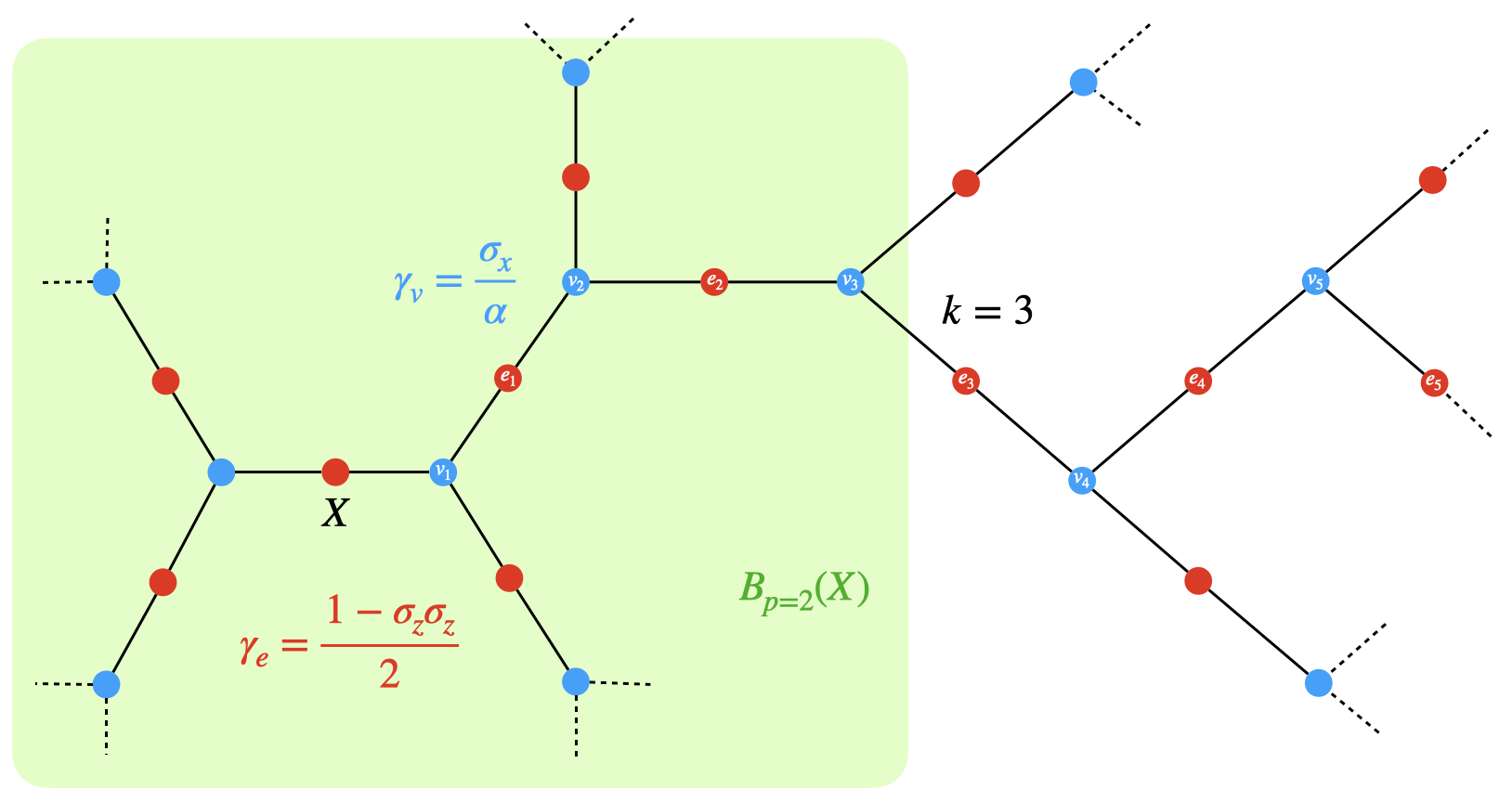}
    \caption{Commutativity graph of the cubic graph that maximizes the LR bound. The shaded area shows an example for $q=2=k-1$.}
    \label{fig:comGraph}
\end{figure}

\begin{equation}
\label{eq:LRbound}
    \lVert \gamma_{X}^A(t) \rVert \leq \left(\frac{2}{\alpha} \right)^{k} \sum_{v_1:\langle {X}v_1 \rangle \in \mathbf{E}} \sum_{e_1:\langle v_1e_1 \rangle \in \mathbf{E}} ... \sum_{e_k:\langle v_{k}e_{k} \rangle \in \mathbf{E}} \int_0^t h_{v_1}(t_1) \int_0^{t_1} h_{e_1}(t_2) ... \int_0^{t_{2k-1}} h_{e_{k}}(t_{2k}) \lVert \gamma_{e_{k}}^A(t_{2k}) \rVert dt_{2k}...dt_2 dt_1
\end{equation}
Now, let us introduce the following nested integral $I_{2k}$ and $I_{2k+1}$ that appears in Equation \ref{eq:LRbound} where we replace each $h_j(t_i)$ by its expression and we pull out the integral the factor $T^{2k}$ and $T^{2k+1}$ respectively so that the integrals depend only on $k$:

\begin{align*}
    I_{2k}&= \int_0^1 1-u_1 \int_0^{u_1} u_2 ... \int_0^{u_{2k-1}} u_{2k} du_{2k}...du_2 du_1 \\
    I_{2k+1}&=\int_0^1 1-u_1 \int_0^{u_1} u_2 ... \int_0^{u_{2k}} 1-u_{2k+1} du_{2k+1}...du_2 du_1
\end{align*} There is no known closed form for these integrals but we can easily have the exact numerical values for at least the first 100 points and we can upper bound it as we show in Appendix \ref{app:integrals}.
We can then write, following Eq.~\ref{eq:LRbound}: 
$$
\lVert \gamma_{X}^A(T) \rVert \leq T^{2k} \left(\frac{2}{\alpha} \right)^{k} I_{2k} \sum_{v_1:\langle {X}v_1 \rangle \in \mathbf{E}} \sum_{e_1:\langle v_1e_1 \rangle \in \mathbf{E}} ... \sum_{e_{k}:\langle v_{k}e_{k} \rangle \in \mathbf{E}}  \max_t \lVert \gamma_{e_{k}}^A(t) \rVert
$$ where $\gamma_{e_{k}}^A(t)$ corresponds to an interaction node of the commutativity graph (i.e. red node) because we are at an even step. We used the fact that $A$ is unitary and the dependence on $B_q(X)$ lies in computing the size of the nested sum. This bound can be improved by applying the update rule of Equation~\ref{eq:normUP} and noticing that the first terms such that $\lVert \gamma_{e_{k}}^A(0) \rVert \neq 0$ only include all paths starting at $X$ and ending in the first red node outside $B_{q=k-1}(X)$ (the green area in Fig. \ref{fig:comGraph}) in $2k$ steps in $\mathbf{G}$. After iterating several times, we get:

\begin{align}
\label{eq:loc_lr}
    \lVert \gamma_{X}^A(T) \rVert  &\leq T^{2k} \left(\frac{2}{\alpha} \right)^k I_{2k}\times \#\{\text{path of length } 2k: Xv_1e_1...v_ke_k \} \lVert \gamma_{e_{k}}^A(0) \rVert \\ \nonumber
    &+ T^{2k+1}\left(\frac{2}{\alpha} \right)^k I_{2k+1}\times \#\{\text{path of length } 2k+1: Xv_1e_1...v_ke_kv_{k+1} \} \lVert \gamma_{v_{k+1}}^A(0) \rVert \\ \nonumber
    &+ T^{2k+2}\left(\frac{2}{\alpha} \right)^{k+1}I_{2k+2} \times \#\{\text{path of length } 2k+2: Xv_1e_1...v_{k+1}e_{k+1} \} \lVert \gamma_{e_{k+1}}^A(0) \rVert\\ \nonumber
    &+ T^{2k+3}\left(\frac{2}{\alpha} \right)^{k+1} I_{2k+3} \times \#\{\text{path of length } 2k+3: Xv_1e_1...v_{k+1}e_{k+1}v_{k+2} \} \lVert \gamma_{v_{k+2}}^A(0) \rVert \\ \nonumber
    &+ T^{2k+4}\left(\frac{2}{\alpha} \right)^{k+2}I_{2k+4}\times \#\{\text{path of length } 2k+4: Xv_1e_1...v_{k+2}e_{k+2} \} \lVert \gamma_{e_{k+2}}^A(0) \rVert\\ \nonumber
    &+ T^{2k+5}\left(\frac{2}{\alpha} \right)^{k+2} I_{2k+5}\times \#\{\text{path of length } 2k+5: Xv_1e_1...v_{k+2}e_{k+2}v_{k+3} \} \lVert \gamma_{v_{k+3}}^A(0) \rVert \\ \nonumber
    &+ T^{2k+6}\left(\frac{2}{\alpha} \right)^{k+3} I_{2k+6}\sum_{v_1:\langle {X}v_1 \rangle \in \mathbf{G}} \sum_{e_1:\langle v_1e_1 \rangle \in \mathbf{G}} ... \sum_{e_{k+2}:\langle v_{k+1}e_{k+2} \rangle \in \mathbf{G}}  \max_t\lVert \gamma_{e_{k+3}}^A(t) \rVert \\ \nonumber
    &= \varepsilon_{loc}(B_{q=k-1}(X),T,\alpha)
\end{align}

We stop at the seventh iteration because numerically it appears that the bound reaches a minimum before increasing again. Each path considered above ends outside $B_q(X)$, because for the others we have $\lVert \gamma_{j}^A(0) \rVert \neq 0$ for $j \in [e_k, v_{k+1},e_{k+1},v_{k+2},e_{k+2},v_{k+3}]$. We also implicitly extend $B_q(X)$ so as to maximize the number of paths in Eq. \ref{eq:loc_lr}. Thanks to the upper bound of the integrals detailed in Appendix \ref{app:integrals}, it is easy to see that the derived bound is decreasing with $k$ and thus with $q$ meaning that we have the following corollary:
\begin{corollary}
\label{cor:mono}
For any $q>0$ and any edge $X$ inside a $d-$regular graph,
    $$\forall j>0, \varepsilon_{loc}(B_{q+j}(X),T,\alpha) \leq \varepsilon_{loc}(B_{q}(X),T,\alpha)$$
the same goes for the global bound as taking the max preserves the inequality:
$$
\forall j>0, \varepsilon(q+j,T,\alpha) \leq \varepsilon(q,T,\alpha)
$$
\end{corollary}

For the local bound, we need to compute for each ball $B_q(X)$ the number of paths. In practice, a subroutine counting the number of paths of a given size is used to compute the local bound. The last term with the multiple sums is counting for $(2d)^{k+3}$ as at each interaction term there are $d$ possible choices of nodes and only 2 at the others. In this subsection we detailed the derivation of the LR local bound Eq. \ref{eq:loc_lr}. In the next subsection, we pursue the derivation to get the global bound by taking the maximal number of paths.

\subsection{Global LR bound}

In this subsection, we use Eq. \ref{eq:loc_lr} to derive the global LR bound. As defined in Section \ref{sec:2}, the global bound is obtained by considering the maximum of the local bound Eq. \ref{eq:loc_lr} over all balls in $\mathcal{B}_q$. To this end, we consider the worst-case scenario, i.e. the ball maximizing the possible number of paths. It is trivial to see that this corresponds to the cycle-free ball (Fig. \ref{fig:comGraph}).

\begin{figure}[ht]
    \centering
    \includegraphics[scale=0.4]{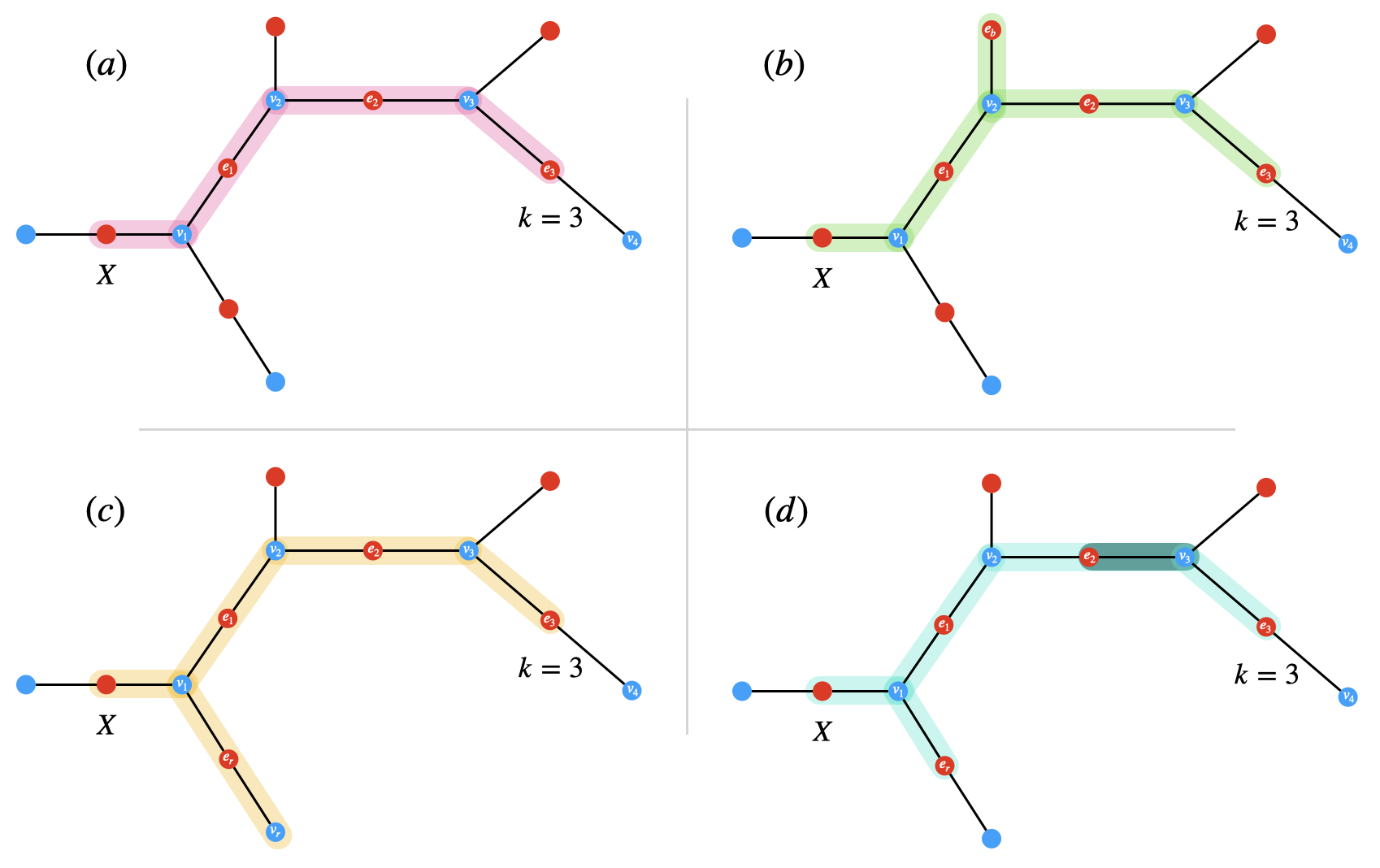}
    \caption{Example of different paths starting at $X$ on the commutativity graph for the global bound with $k=3$. $(a)$ a simple path $({\color{red}X}{\color{blue}v_1}{\color{red}e_1}{\color{blue}v_2}{\color{red}e_2}{\color{blue}v_3}{\color{red}e_3})$ in pink of length $2k$, $(b)$ a green path $({\color{red}X}{\color{blue}v_1}{\color{red}e_1}{\color{blue}v_2}{\color{red}e_b}{\color{blue}v_2}{\color{red}e_2}{\color{blue}v_3}{\color{red}e_3})$ of length $2k+2$ with a back-and-forth on one edge, $(c)$ in orange a path $({\color{red}X}{\color{blue}v_1}{\color{red}e_r}{\color{blue}v_r}{\color{red}e_r}{\color{blue}v_1}{\color{red}e_1}{\color{blue}v_2}{\color{red}e_2}{\color{blue}v_3}{\color{red}e_3})$ of length $2k+4$ with a back-and-forth on a branch of two edges and $(d)$ in blue a path $({\color{red}X}{\color{blue}v_1}{\color{red}e_r}{\color{blue}v_1}{\color{red}e_1}{\color{blue}v_2}{\color{red}e_2}{\color{blue}v_3}{\color{red}e_2}{\color{blue}v_3}{\color{red}e_3})$ of length $2k+4$ with two back-and-forth on two edges at distance at most one from the simple path.}
    \label{fig:paths}
\end{figure}

In this cycle-free ball, we can count the number of paths corresponding to each term in LR bound's equation. In Fig. \ref{fig:paths}, we depict example paths for each of the necessary cases that we detail below. 

For the first two terms, only direct simple paths reach the outside of $B_q(X)$, and there are $2(d-1)^k$ of them. The factor 2 comes from the initial choice at node $X$, you can go either left or right on Fig. \ref{fig:comGraph}. Once the side has been chosen, at each blue node (node $v_r$), there are $d-1$ possibilities, as the path can not go backwards by definition of direct single paths. In a path from $X$ to the first node outside $B_{q=k-1}(X)$, i.e. of length $2k$, there are $k$ blue nodes, bringing the total number of direct simple paths to $2(d-1)^k$ (path Fig. \ref{fig:paths} $(a)$). The same number of paths is found for direct paths of length $2k+1$.

Then, for the third and fourth terms, we can distinguish simple direct paths that go one step further in $\mathbf{G}$, i.e. of length $2k+2$ and $2k+3$ respectively, from non-direct paths, i.e. passing several times through the same node or edge. 
For the third term counting paths of length $2k+2$, similarly to above, there are $2(d-1)^{k+1}$ direct paths. For the non-direct ones, we need to count every edge that can be used at least twice in the path (path Fig. \ref{fig:paths} $(b)$. At first, there are the two edges that start from $X$, then $(d-1)$ at each blue node on the path and $+1$ at each red node, making a total of $2+((d-1)+1)*k$ edges that can be used twice in the $2(d-1)^k$ possible paths. Therefore, the total number of paths in the third term is $2(d-1)^{k+1}+(2+dk)*2(d-1)^k$. For the fourth term, we use similar reasoning to arrive at $2(d-1)^{k+1}+(dk+2+d-1)*2(d-1)^k$ paths. 

Let us see how the last two terms are derived. For the fifth term, we need to count all paths of length $2k+4$ that lead to a red node outside the shaded area. There are three types of path: direct paths up to $e_{k+2}$ counting for $2(d-1)^{k+2}$, those with exactly one edge taking two or three times, i.e. reaching $e_{k+1}$ counting for $(d(k+1)+2)*2(d-1)^{k+1}$ and those reaching $e_{k}$ counting for $\left [ (dk+2) + \binom{dk+2}{2} + 2(d-1)+k \right ]*2(d-1)^k $. In the latter, a distinction is made: either an edge is used 4 or 5 times, or 2 different edges at a distance at most one from the direct path can be used 2 or 3 times (path Fig. \ref{fig:paths} $(d)$), or finally choose a branch of length 2 far from the direct path (path Fig. \ref{fig:paths} $(c)$). Similar reasoning is used for the sixth term.

We can then substitute the path counting in Eq. (\ref{eq:loc_lr}) to derive the following closed-form:

\begin{align}
\label{eq:LR7t}
    \lVert \gamma_{X}^A(t) \rVert &\leq T^{2k}\left(\frac{2}{\alpha} \right)^k I_{2k} \times 2(d-1)^k  +T^{2k+1}\left(\frac{2}{\alpha} \right)^{k+1} I_{2k+1} \times 2(d-1)^k  \\ \nonumber 
    &+ T^{2k+2}\left(\frac{2}{\alpha} \right)^{k+1}I_{2k+2} \times 2(d-1)^k \left [ d(k+1)+1 \right] + T^{2k+3}\left(\frac{2}{\alpha} \right)^{k+2}I_{2k+3} \times 2(d-1)^k \left [ d(k+2) \right]\\ \nonumber
    &+ T^{2k+4}\left(\frac{2}{\alpha} \right)^{k+2}I_{2k+4} \times 2(d-1)^k\left [ d^2\frac{(k+1)^2+3}{2}+d\frac{3k+2}{2}+k \right]\\ \nonumber
    &+ T^{2k+5}\left(\frac{2}{\alpha} \right)^{k+3}I_{2k+5} \times 2(d-1)^k\left [ d^2\frac{(k+2)^2+3}{2}+d\frac{k+1}{2}+k-1 \right]\\ \nonumber
    &+ T^{2k+6}\left(\frac{2}{\alpha} \right)^{k+3} I_{2k+6} \times (2d)^{k+3}= \varepsilon(q=k-1,T,\alpha)
\end{align}

In this subsection, we developed the proof of our LR bound for any $d-$regular graph on which we want to solve MaxCut with a quantum annealing process. This bound achieves the best numerical value compared to the state-of-the-art of LR bounds. This is due to the fact that we have finely evaluated the nested integral with the standard schedule and used the commutativity graph of the Hamiltonian to tighten the bound. 
Here the free parameter $\alpha$ plays an important role: optimizing over its value will allow us to control the tightness of the bound (\ref{eq:LR7t}). This point is further discussed in Section \ref{sec:discussion}.
In the next subsection, we apply the derived bounds (global and local) to obtain a numerical value of the approximation ratio for a 1-local analysis of QA.

\subsection{Application to approximation ratio of MaxCut}
\label{ssec:app}
 In this subsection, we use the previously derived LR bounds to determine the approximation ratio of MaxCut over cubic graph with QA analyzed as a $1-$local algorithm. The proof of the Theorem \ref{thm:main} proceeds with step 3, 4 and eventually 5, as illustrated in the overview of Fig. \ref{fig:overview}. We will use the Eq. \ref{eq:minrho} to derive the approximation ratio with the $1-$local analysis. 

For this purpose, after rigorous errors and trials, we set specific values $T=3.33$, $\alpha=1.53$ and $q=3$, that establish the global bound $\varepsilon(q,T,\alpha)$. In order to compute the required minimums of Eq. \ref{eq:minBq}, $\langle O_X \rangle_{\mathcal{B}_{3,i}}^*$, we need to enumerate all balls in $\mathcal{B}_3$ and all cubic graphs in $\mathcal{B}_2$. We follow a smart hash-based iterative algorithm detailed in Appendix \ref{app:enum}. The algorithm generated 930449 balls. Employing the AnalogQPU simulator of Eviden Qaptiva see Appendix \ref{app:enum}), we solve the Schrödinger equation to get the final state $|\psi^{B_3(X)}(T,\alpha)\rangle $ as described in the paragraph “Parametrized QA” of Section \ref{sec:2}. This allows to explicitly evaluate the value of $\langle O_X \rangle_{B_3(X)}$ for each ball in $\mathcal{B}_3$. We subtract the value of the global LR bound to the expected edge energy for each ball. To narrow down our selection, we retain only those balls for which $|\langle O_X \rangle_{B_3(X)}-\varepsilon(q,T,\alpha)|<0.7020$. This initial step corresponds to step 3 of Fig. \ref{fig:overview} and leads to the values for $\langle O_X \rangle_{\mathcal{B}_{3,1}}^*=0.5502$ and $\langle O_X \rangle_{\mathcal{B}_{3,2}}^*=0.6265$. These values satisfy the condition (see Appendix \ref{app:min}) where the ratio reduces to $\rho_{MC} \geq \langle O_X \rangle_{\mathcal{B}_{3,3}}^*$. In other words, the critical balls correspond to configurations $\Omega_3$ (see Fig.~\ref{fig:omegas}), as it usually happens in QA and QAOA algorithms for MaxCut~\cite{farhi2014quantum,braida2022constant}. Consequently, our goal is to maximize this minimum \footnote{The Open Source code for reproducibility of this work is made available to readers on GitHub \href{https://github.com/Arts-Braido/LR-bound-for-approximation}{https://github.com/Arts-Braido/LR-bound-for-approximation}. }.
 
\begin{figure}[ht]
    \centering
    \includegraphics[width=1\textwidth]{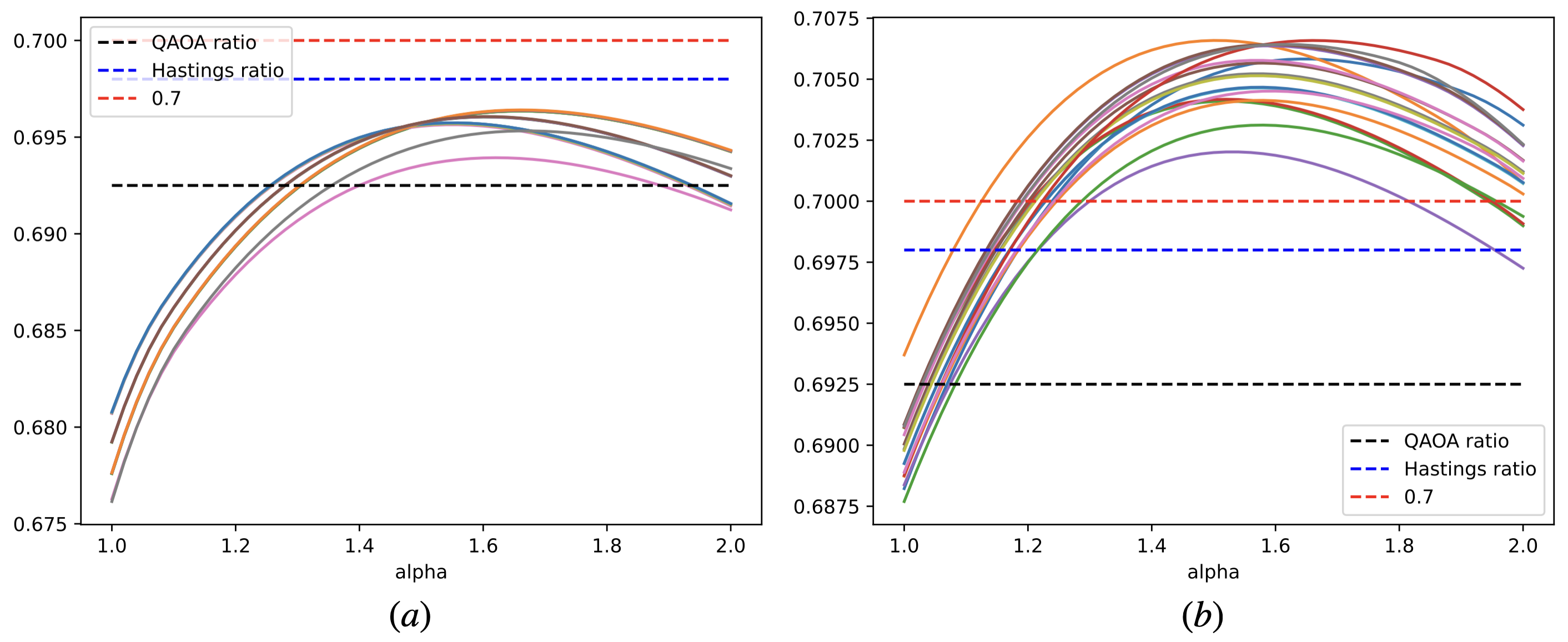} 
    \caption{Evolution of $(a)$ $\max_T \left (\langle O_X \rangle_{B_3(X)}-\varepsilon(3,T,\alpha) \right )$ and $(b)$ $\max_T \left (\langle O_X \rangle_{B_3(X)}-\varepsilon_{loc}(B_3(X),T,\alpha) \right )$ against $\alpha$ for the 18 worst balls for which it goes under 0.7020 with the global bound.}
    \label{fig:worstloc}
\end{figure}

We are left with 7071 balls $B_3(X)$ with configuration $\Omega_3$ at distance 1 around $X$, to which we apply the local bound. To compute the local bound we have access to a path counting algorithm as there is no closed form like the global bound. To find the maximum over parameters $T$ and $\alpha$, on Fig. \ref{fig:worstloc}, we plot the evolution of (a) $\max_T \left (\langle O_X \rangle_{B_3(X)}-\varepsilon(3,T,\alpha) \right )$ and (b) $\max_T \left (\langle O_X \rangle_{B_3(X)}-\varepsilon_{loc}(B_3(X),T,\alpha) \right )$ against $\alpha$ for the 18 worst balls $B_3(X)$ according to the global and local bounds respectively. 

\begin{figure}[ht]
    \centering
    \includegraphics[scale=0.3]{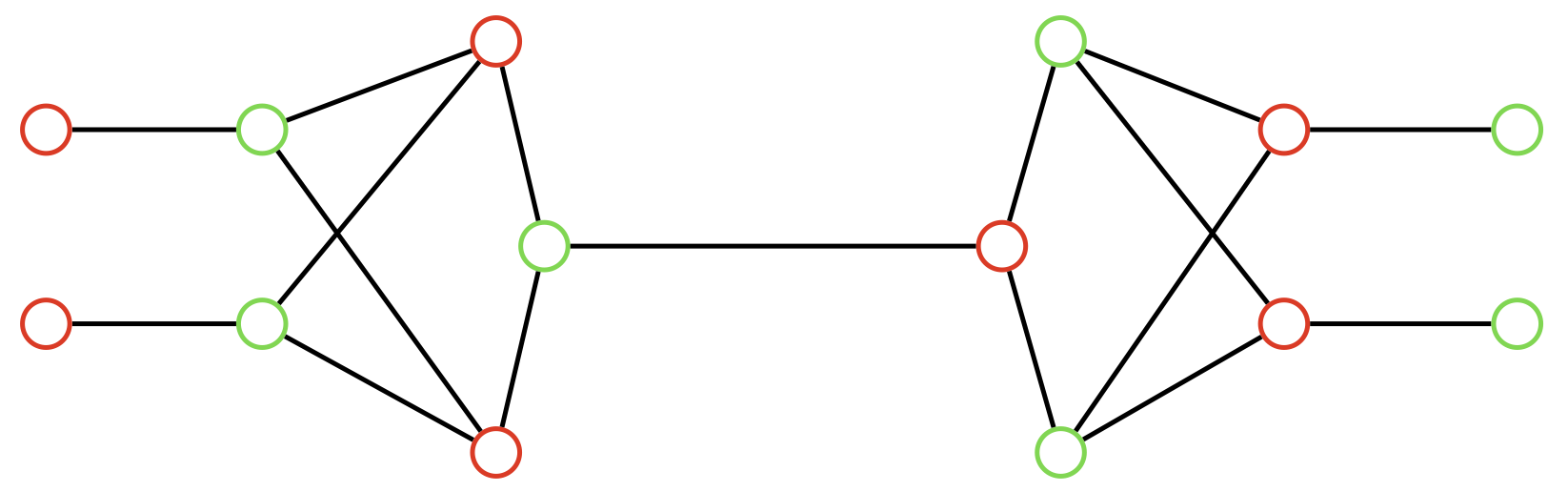}
    \caption{Ball $B_3(X)$ achieving the minimum $\langle O_X \rangle_{\mathcal{B}_{3,3}}^*=0.70208...$}
    \label{fig:worst_ball}
\end{figure}

The analysis reveals that  around $\alpha=1.5$ all these balls surpass the threshold of 0.7020, with the worst ball depicted in Fig. \ref{fig:worst_ball}. This plot finally fixes the value of $\langle O_X \rangle_{\mathcal{B}_{3,3}}^*=0.70208...$ which proves Theorem~\ref{thm:main}. 

To sum up, the constant-time analysis of Quantum Annealing (QA) for MaxCut over cubic graphs, analysed as a 1-local algorithm, achieves an approximation ratio exceeding 0.7020. This result goes beyond any known ratio of 1-local algorithms, whether quantum or classical. 

\section{Discussion}
\label{sec:discussion}

\begin{figure}[ht]
    \centering
    \includegraphics[scale=0.6]{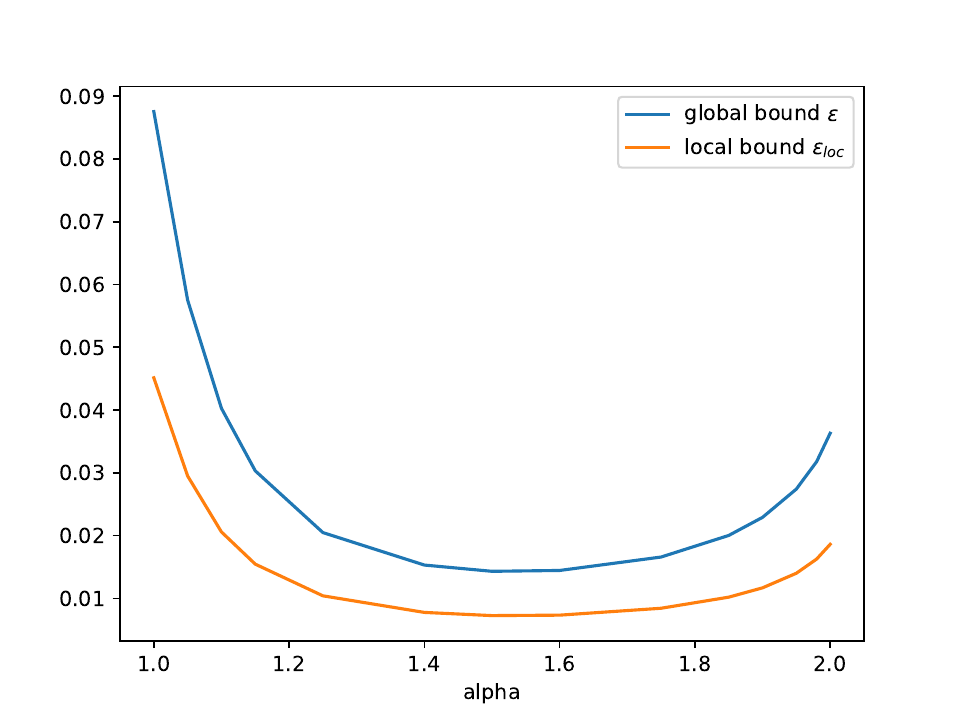}
    \caption{Evolution of $\varepsilon_{loc}(g,T,\alpha)$ and $\varepsilon(q=3,T,\alpha)$ against $\alpha$ for pairs $(T,\alpha)$ for which $\langle O_X \rangle_{g} = 0.7092 $ and $g$ being the ball of Fig. \ref{fig:worst_ball}.}
    \label{fig:tightness}
\end{figure}

In this section, we discuss the result and the tightness of Theorem~\ref{thm:main}. In addition to the use of the commutativity graph, the exact value of the nested integrals alone would not have brought the ratio above the targeted numerical values of QAOA and Hastings local algorithm as we see on Fig. \ref{fig:worstloc} (b) at $\alpha=1$. The introduction of the “hyperparameter” $\alpha$ significantly enhances the precision of the analysis. To attest this point, on Fig. \ref{fig:tightness}, we plot the evolution of both the local $\varepsilon_{loc}(g,T,\alpha)$ and global $\varepsilon(q,T,\alpha)$ bounds against $\alpha$ for pairs $(T,\alpha)$ for which $\langle O_X \rangle_{g} = 0.7092 $ and $g$ denotes the ball of Fig. \ref{fig:worst_ball}. The value of 0.7092 is totally arbitrary and similar plots are achieved with different values. So we clearly see on Fig. \ref{fig:tightness} that the LR bound is minimal around $\alpha=1.5$ which means that the analysis of QA is tighter around this point. 

We would like to draw readers' attention to the fact that both the Hastings algorithm and QAOA also include one or two hyperparameters for obtaining their best ratio value. In this sense, our parameterized QA analysis is nothing more complex. However, although these two other algorithms produce a tight ratio value, QA's is not tight simply because it is impossible to construct a graph such that every edge $X$ has the $g$ from Fig. \ref{fig:worst_ball} as its $B_3(X)$.

On the LR bound itself, we can see from Fig. \ref{fig:worstloc} $(a)$ that for many of the balls, the curves of the edge energy are indistinguishable. Indeed, at the pair $(T,\alpha)$ looked at, there is no visible difference in the value of the average edge energy. Even two balls in $\mathcal{B}_2$ can have the same trajectory. Informally, this suggests that LR bound is not yet tight, as the layer at distance 3 from the edge $X$ does not impact the edge energy. As noted in \cite{braida2022constant}, neglecting the initial state greatly affects the accuracy of the bound. 
In order to strongly improve the analysis it would be crucial to be able to take into consideration the initial state; unfortunately, at this stage we lack of mathematical tools for this purpose. Another lead is indicated in \cite{chen2023speed}, in path counting. In fact, it seems that only direct paths contribute to LR bound. We explored this idea and the bound would reach 0.7041. However the result of~\cite{chen2023speed} cannot be adapted as is to our framework, so we cannot claim this approximation ratio at this stage. 

\paragraph{Toward better ratios: }An important field in the improvement of QA's performance is the optimization of the schedule. Indeed, all the construction above works for a linear interpolation with no specific optimization but one can look at the Hamiltonian 
$$
H(t,G)=(1-f(t/T))H_0(\alpha)+f(t/T)H_1^G
$$
with $f(t)$ such that $f(0)=0$ and $f(1)=1$. It is important to note that it also modifies the LR bound in the nested integrals. In commutativity graph notation, we have the $h_v(t)=1-f(t/T)$ and $h_e(t)=f(t/T)$. The challenge of this optimization is that one has to evaluate the energy on each of the 930449 balls for each tested schedule to formally prove an improvement of the ratio. However, a good insight can be reached by looking only at the 123 balls in $\mathcal{B}_2$. For instance, we tried few cubic functions to already see that with $f(s)=3.2s^3-4.8s^2+2.6s$, we should obtain a ratio above 0.7165 at $T=2.77$ and $\alpha=1.6$.

\paragraph{Directions for generalizing the construction.} 
We derived LR-type bounds applied to MaxCut on cubic graphs. We believe that our tools can be extended in several directions.
\begin{enumerate}
    \item This work can be directly adapted to look for a other $d-$regular graphs. A formal derivation of this bound for any $d-$regular graphs requires to enumerate all the balls in $\mathcal{B}_3$ that can be completed in a $d-$regular graphs. The code can be easily edited for this purpose. However large balls are certainly too big to solve Schrodinger equation for $d \geq 4$. It is still possible to get an intuition extrapolating the worst ball for $d=3$ (Fig. \ref{fig:worst_ball}). For example, the $1-$local approximation ratio for MaxCut over 4-regular graphs may be close to 0.67. 
    \item For $p-$local analysis of QA with $p\geq 2$, the method developed here runs short as the time at which the best approximation ratio is achieved is certainly too large for the LR bound at $q=3$. For a $p=2-$local analysis, our estimation gives that we would need to go up to $q=5$ to achieve sufficiently small LR bound at time $T\simeq 6.1$, time for which the expected edge energy value seems to maximize. Nevertheless, by extrapolating at $p=2$ the worst-case balls for $p=1$ and by numerical experiments on these cases, we believe that the approximation ratio for MaxCut over cubic graphs is close to 0.77. 
    \item As discussed in the previous paragraph, the schedule can also be changed, the main work remains in the computation of the nested integrals of the schedule. Analytical bounds on these integrals are certainly too difficult to derive, but only numerical values are required to prove the bound. For any polynomial schedule, those integrals are easy to evaluate.
    \item This construction can be applied to other combinatorial graph problems. For instance, in \cite{braida2022constant}, the authors applied a similar analysis to the Maximum Independent Set problem over cubic-graphs. More work is needed to adapt derive an ad hoc analytical formula for LR bound for this new problem Hamiltonian.
\end{enumerate}

\subsection*{Conclusion} 
To conclude, in this work we developed a much tighter Lieb-Robinson bound compared to \cite{Haah_2021,chen2023speed} by carefully manipulating the commutativity graph and the nested integral of the QA schedule. 
Despite the continuous aspect of QA, we defined the notion of $p-$local analysis of the metaheuristic by approximating the full algorithm using its restriction to bounded radius subgraphs.
Our 1-local analysis of QA allows us to  analytically compare its performances with the performances of single-layer QAOA for MaxCut over cubic graphs. The tightness of the LR bound we have derived enables us to reduce the exhaustive numerical simulation to a tractable task that can be completed in a few weeks. Finally, we introduced a parameter in the standard QA, enabling us to optimize the value of the ratio obtained and thus pass the 0.7 mark with a ratio going beyond 0.7020. This puts us ahead of single-layer QAOA and Hastings' $1-$local algorithm for MaxCut over cubic graphs. The comparison has its limits, as the process we are studying is continuous and not intrinsically local, unlike the two algorithms mentioned. This work should be seen as a step forward in the study of quantum annealing, bringing more analytical tools to assess its algorithmic performances. 

For future work, LR bound improvement can be tackled by considering the initial state information, which is not arbitrary. In \cite{braida2022constant}, the authors show that there is a considerable lost in tightness of the bound by neglecting the initial state. Also, as mentioned above, a slight improvement can be made to path counting. Then, 
our tight numerical result might be applied for a practical implementation of some Hamiltonian simulation schemes, e.g.~\cite{Haah_2021}.

\section*{Acknowledgment}
This work is part of HQI initiative \href{www.hqi.fr}{(www.hqi.fr)} and is supported by France 2030 under the French National Research Agency award number “ANR-22-PNCQ-0002”. The funder played no role in study design, data collection, analysis and interpretation of data, or the writing of this manuscript

\appendix
\section{Minimization of Equation \eqref{eq:min1loc}}
\label{app:min}
Let us study the following function with two variables:
$$\forall x\geq 0,y \geq 0 | 4x+3y \leq1, f(x,y)=\frac{ax+b(4x+3y)+c(\frac{3}{2}-5x-3y)}{\frac{3}{2}-x-y}$$ 
where $a=\langle O_X \rangle_{\mathcal{G}}^{\Omega_1}$, $b=\langle O_X \rangle_{\mathcal{G}}^{\Omega_2}$ and $c=\langle O_X \rangle_{\mathcal{G}}^{\Omega_3}$. Empirically we suppose that $a\leq b\leq c$ and we will see later that this assumption is verified. Let $x,y$ be positive number such that $4x+3y \leq 1$. Then we have that $x+y\leq \frac{1}{3}$ and $x\leq \frac{1}{4}$. The function $f$ can be rewritten in three parts:
$$
f(x,y)=\frac{\frac{3}{2}c}{\frac{3}{2}-(x+y)}-\frac{3(c-b)(x+y)}{\frac{3}{2}-(x+y)} - \frac{(2c-a-b)x}{\frac{3}{2}-(x+y)} 
$$
If $x=y=0$, then $f(0,0)=c$. Let us see the condition on which $f$ can only increase if $x+y>0$. The first term can increase at most by $c-c\frac{\frac{3}{2}}{\frac{3}{2}-\frac{1}{3}}=\frac{2}{7}c$. The second term can decrease at most by $\frac{c-b}{\frac{3}{2}-\frac{1}{3}}=(c-b)\frac{6}{7}$. The last term can decrease by at most $\frac{\frac{1}{4}(2c-a-b)}{\frac{3}{2}-\frac{1}{3}}=(2c-a-b)\frac{3}{14}$. Thus $f(0,0)$ is the minimum if $\frac{2}{7}c\geq (c-b)\frac{6}{7}+(2c-a-b)\frac{3}{14}$. Therefore, we can derive the following condition to satisfy to have $f(0,0)$ as the minimum:
\begin{align*}
    c &\geq 3(c-b)+\frac{3}{4}(2c-a-b) \\
    \Rightarrow  3.5c &\leq 3.75b+0.75a
\end{align*}
For example with $a\geq 0.5$, $b\geq 0.57$ and  $c\leq 0.71$ the assumption and the condition are satisfied. 

\section{Proof of Equation \eqref{eq:normUP}}
\label{app:proof}
We are working with a Hamiltonian of general form $H(t) = \sum_{j \in V(\mathbf{G})} h_j(t) \gamma_j$ and for any unitary $A$ supported on $S$ (outside a certain region around one node $X$), we want to show Equation \eqref{eq:normUP} in the time-dependent regime. We follow exactly the same steps of \cite{Wang_2020} but with function $h_j$ that depends on the time. First let us look at the derivative of $\gamma_{X}^A(t)$:
$$
\frac{d[\gamma_{X}(t),A]}{dt} = -i \sum_{v:\langle Xv \rangle \in \mathbf{G}} h_{v}(t) [(U_T^G)^\dagger [\gamma_{X}, \gamma_{v}] U_T^G,A]
$$
Then we define $\tau_A(t)=\hat{U}^\dagger \gamma_{X}^A(t)\hat{U} $ where $\hat{U}$ is the unitary that is solution of $i\frac{d\hat{U}}{dt} = -\sum_{v:\langle Xv \rangle \in \mathbf{G}} h_{v}(t) \gamma_{v}(t) \hat{U}$. That way, we have $ \lVert \tau_A(T) \rVert= \lVert \gamma_{X}^A(T) \rVert$ and the derivative of $\tau_A$ is given by:
$$
\dot{\tau_A}(t)=-i  \sum_{v:\langle Xv \rangle \in \mathbf{G}} h_{v}(t) \hat{U}^\dagger \left [ \gamma_{X}(t),[\gamma_{v}(t),A]\right] \hat{U}
$$
Now we can proceed as follow:
\begin{align*}
    \lVert \gamma_{X}^A(T) \rVert - \lVert \gamma_{X}^A(0) \rVert &= \lVert \tau_A(T) \rVert-\lVert \tau_A(0) \rVert\\
    &\leq \int_0^T \| \dot{\tau_A}(t') \| dt'\\
    &\leq \sum_{v:\langle Xv \rangle \in \mathbf{G}}\int_0^T h_{v}(t) \left \lVert \left [ \gamma_{X}(t),[\gamma_{v}(t),A] \right ] \right \rVert dt
\end{align*} \qed

\section{Nested integrals}
\label{app:integrals}
In this appendix, we detail the computation of the nested integrals that play an important role in the LR-bound. To tackle this derivation, we introduce for all $k\in \mathbf{N}^*$: $$I_{2k}(x)= \int_0^x 1-u_1 \int_0^{u_1} u_2 ... \int_0^{u_{2k-1}} u_{2k} du_{2k}...du_2 du_1 $$ and $$I_{2k+1}(x)= \int_0^x 1-u_1 \int_0^{u_1} u_2 ... \int_0^{u_{2k}} 1-u_{2k+1} du_{2k+1}...du_2 du_1 $$ polynomials in $x$ define in $[-1,1]$. The goal is to compute these polynomials at $x=1$, which is nothing else than the sum of the polynomial coefficients.  

\paragraph{Even case $I_{2k}(x)$:}The highest degree of this polynomial is $4k$ as there are $2k$ integrals and there is always a way to choose a $u_j$ to integrate so the $2k$ $u_j$ bring the total degree to $4k$. The highest order coefficient is straightforward $\frac{(-1)^k}{2^{2k}(2k)!}$. The least order term is the one where we choose the minimum number of $u_j$ to integrate which is $k$. This observation brings the least order term to be of degree $3k$ with coefficient $\prod_{l=2}^k \frac{1}{3l(3l-1)}$. In general, there exist positive coefficients $a_j(k)$ to express $I_{2k}(x)$ as a polynomial:
\begin{align}
    I_{2k}(x) &= \sum_{j=0}^k a_j(k) (-1)^j x^{3k+j} \\
    &= x^{3k} \sum_{j=0}^k a_j(k) (-x)^j
\end{align}
and we define $a_0(0)=1$. With these notations, the quantity of interest is $I_{2k}(1)=\sum_{j=0}^k (-1)^j a_j(k) $. To find a recurrence relation first notice that $I_{2k}(x) = \int_0^x 1-u_1 \int_0^{u_1} u_2 I_{2k-2}(u_2) du_2 du_1$ and develop:

\begin{align}
    I_{2k}(x) 
    &= \int_0^x 1-u_1 \int_0^{u_1} u_2 * I_{2k-2}(u2) du_2 du_1 \nonumber \\
    &= \sum_{j=0}^{k-1} a_j(k-1) (-1)^j \int_0^x 1-u_1 \int_0^{u_1} u_2^{3k+j-2} du_2 du_1 \nonumber \\
    &= \sum_{j=0}^{k-1} a_j(k-1) (-1)^j \int_0^x (1-u_1) \frac{u_1^{3k+j-1}}{3k+j-1} du_1 \nonumber \\
    &= \sum_{j=0}^{k-1} a_j(k-1) (-1)^j \left [\frac{x^{3k+j}}{(3k+j-1)(3k+j)} - \frac{x^{3k+j+1}}{(3k+j-1)(3k+j+1)} \right ] \nonumber \\
    &= x^{3k} \sum_{j=0}^{k-1} a_j(k-1) (-1)^j \left [\frac{x^{j}}{(3k+j-1)(3k+j)} - \frac{x^{j+1}}{(3k+j-1)(3k+j+1)} \right ] \nonumber \\
    &= x^{3k} \sum_{j=0}^{k-1} a_j(k-1) (-1)^j \frac{x^{j}}{(3k+j-1)(3k+j)} - x^{3k} \sum_{j=1}^{k} a_{j-1}(k-1) (-1)^{j-1}\frac{x^{j}}{(3k+j-2)(3k+j)} \nonumber \\
    &= \frac{a_0(k-1)}{3k(3k-1)}x^{3k} + x^{3k} \sum_{j=1}^{k-1} \frac{(3k+j-2)a_j(k-1) + (3k+j-1)a_{j-1}(k-1)}{(3k+j-2)(3k+j-1)(3k+j)} (-x)^{j} + \frac{(-1)^k a_{k-1}(k-1)}{4k(4k-2)}x^{4k}
\end{align}
In the last line we can identify the $a_j(k)$ coefficients:
\begin{align}
    a_0(k) &= \frac{a_0(k-1)}{3k(3k-1)} \\
    a_j(k) &= \frac{a_j(k-1)}{(3k+j-1)(3k+j)}+\frac{a_{j-1}(k-1)}{(3k+j-2)(3k+j)} \quad \text{for } j \in [1,...,k-1] \\
    a_k(k) &= \frac{ a_{k-1}(k-1)}{4k(4k-2)}
\end{align}
We recover the higher and least order coefficients mentioned above. So we can try to compute $I_{2k}(1)$:
\begin{align}
    I_{2k}(1) 
    &= \sum_{j=0}^k (-1)^j a_j(k) \nonumber \\
    &= \frac{a_0(k-1)}{3k(3k-1)} + \sum_{j=1}^{k-1} (-1)^j \left [ \frac{a_j(k-1)}{(3k+j-1)(3k+j)}+\frac{a_{j-1}(k-1)}{(3k+j-2)(3k+j)} \right] + (-1)^k \frac{ a_{k-1}(k-1)}{4k(4k-2)} \nonumber \\
    &= \frac{a_0(k-1)}{3k(3k-1)} + \sum_{j=1}^{k-1} (-1)^j \frac{a_j(k-1)}{(3k+j-1)(3k+j)}+\sum_{j=0}^{k-2} (-1)^{j+1} \frac{a_{j}(k-1)}{(3k+j-1)(3k+j+1)} + (-1)^k \frac{ a_{k-1}(k-1)}{4k(4k-2)} \nonumber \\
    &= \sum_{j=0}^{k-1} \frac{(-1)^j a_j(k-1)}{(3k+j-1)(3k+j)(3k+j+1)} 
\end{align}
Computing recursively the coefficients $(a_j(k))_j$ allows us to evaluate the nested integrals of interest. At the end of this appendix, we develop the numerical analysis of it.

We can still have a loose upper bound (sufficient for Corollary 3.2) because the $a_j(k)$ are positives, by looking at $I_{2k}(-1)$:
\begin{align}
    I_{2k}(-1) 
    &= \sum_{j=0}^k a_j(k) \nonumber \\
    &= \frac{a_0(k-1)}{3k(3k-1)} + \sum_{j=1}^{k-1} \left [ \frac{a_j(k-1)}{(3k+j-1)(3k+j)}+\frac{a_{j-1}(k-1)}{(3k+j-2)(3k+j)} \right] + \frac{ a_{k-1}(k-1)}{4k(4k-2)} \nonumber \\
    &= \frac{a_0(k-1)}{3k(3k-1)} + \sum_{j=1}^{k-1} \frac{a_j(k-1)}{(3k+j-1)(3k+j)}+\sum_{j=0}^{k-2} \frac{a_{j}(k-1)}{(3k+j-1)(3k+j+1)} + \frac{ a_{k-1}(k-1)}{4k(4k-2)} \nonumber \\
    &= \sum_{j=0}^{k-1} \frac{6k+2j+1}{(3k+j-1)(3k+j)(3k+j+1)} a_j(k-1) \\
    &\leq \frac{6k+1}{(3k-1)3k(3k+1)} I_{2k-2}(-1) \nonumber \\
    & \leq I_2(-1) \prod_{l=2}^k \frac{6l+1}{(3l-1)3l(3l+1)} \nonumber \\
    &\leq \frac{6^{k+1} k!}{(3k+1)!}
\end{align}

\paragraph{Odd case $I_{2k+1}(x)$:} The higher degree of this polynomial is $4k+2$ as there are $2k+1$ integrals and there is a way to always choose a $u_j$ to integrate so the $2k+1$ $u_j$ bring the total degree to $4k+2$. The least order term is the one where we choose the minimum number of $u_j$ to integrate which is $k$. This observation brings the least order term to be of degree $3k+1$. In general, there exist positive coefficients $b_j(k+1)$ to express $I_{2k+1}(x)$ as a polynomial:
\begin{align}
    I_{2k+1}(x) &= \sum_{j=0}^{k+1} b_j(k+1) (-1)^j x^{3k+j+1} \\
    &= x^{3k+1} \sum_{j=0}^{k+1} b_j(k+1) (-x)^j
\end{align}

\begin{align}
    I_{2k+1}(x) 
    &= \int_0^x 1-u_1 \int_0^{u_1} u_2 * I_{2k-1}(u_2) du_2 du_1 \nonumber \\
    &= ... \nonumber \\
    &= \frac{b_0(k)}{3k(3k+1)}x^{3k+1} + x^{3k+1} \sum_{j=1}^{k} \frac{(3k+j-1)b_j(k) + (3k+j)b_{j-1}(k)}{(3k+j+1)(3k+j-1)(3k+j)} (-x)^{j} + \frac{(-1)^{k+1} b_{k}(k)}{4k(4k+2)}x^{4k+2}
\end{align}
We can identify the $b_j(k)$ coefficients:
\begin{align}
    b_0(k+1) &= \frac{b_0(k)}{3k(3k+1)} \\
    b_j(k+1) &= \frac{b_j(k)}{(3k+j+1)(3k+j)} + \frac{b_{j-1}(k)}{(3k+j+1)(3k+j-1)} \quad \text{for } j \in [1,...,k] \\
    b_{k+1}(k+1) &= \frac{b_{k}(k)}{4k(4k+2)}
\end{align}
Those coefficients allows us to compute the exact value of the nested integral.
Similarly to the even case, we can upper bound the integral of interest like this:
\begin{align}
    I_{2k+1}(1) 
    &= \sum_{j=0}^{k+1} (-1)^j b_j(k+1) \nonumber \\
    &= \sum_{j=0}^{k} \frac{(-1)^j b_j(k)}{(3k+j)(3k+j+1)(3k+j+2)} \\
    I_{2k+1}(1) &\leq I_{2k+1}(-1) \leq \frac{6^{k+2} (k+1)!}{(3k+2)!}
\end{align}

\begin{figure}[ht]
    \centering
    \includegraphics[scale=0.5]{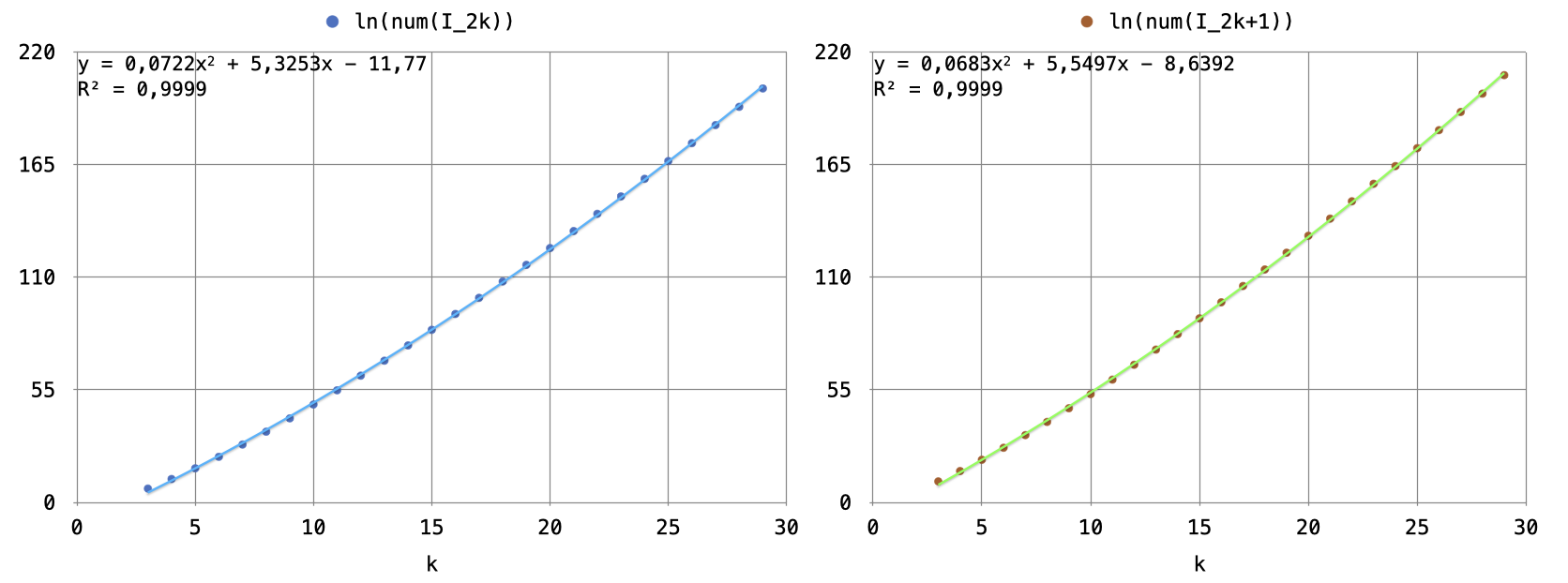}
    \caption{Fit curves for the numerator logarithm of the nested integral for even (left) and odd (right) iteration step.}
    \label{fig:nestedInt}
\end{figure}

In practice, a numerical study is enough as we are interested in the numerical value. Using the derived coefficients above, it appears that it follows something like $I_{2k} \simeq \frac{\text{num}(I_{2k})}{(4k)!}$ and $I_{2k+1} \simeq \frac{\text{num}(I_{2k+1})}{(4k+2)!}$ where both num$(I_{2k+1})$ and num$(I_{2k})$ grow like $e^{\mathcal{O}(k^2)}$. A numerical fit gives a tendency function with an almost 1 correlation factor, see Fig. \ref{fig:nestedInt}.

\section{Balls enumeration and numerical simulation }
\label{app:enum}
In this appendix, we detail the specificity of our balls enumeration algorithm. We use the same idea as that developed in \cite{MaxCutp2}, i.e. an iterative algorithm. We first enumerate all the balls in $\mathcal{B}_1$ which are those on Fig. 3. Then, starting from $\mathcal{B}_{p-1}$, we enumerate $\mathcal{B}_p$ by completing the nodes on the boundary whose degree is less than $d$ of the balls in $\mathcal{B}_{p-1}$. There are different ways of completing a node: either we link it to another existing node of degree less than $d$, or we create a new node and link it. By doing this on all nodes of degree less than $d$ of a ball in $\mathcal{B}_{p-1}$, you get a ball in $\mathcal{B}_p$. You then apply an isomorphism test to check that the algorithm has not already generated isomorphic balls. The isomorphism test must also take into account the fact that there is a “marked” edge $X$ in the balls $B_p(X)$. The edge\_match argument in the Networkx isomorphism test takes this into consideration.

A naive approach could take more than one year to enumerate all the balls in $\mathcal{B}_3$. The isomorphism test is quite expensive as we do not have an efficient algorithm for it. Our idea is to hash the generated balls into a dictionary. The prerequisite for the hash function $h$ to be useful is that for any two graphs $G_1$ and $G_2$, $h(G_1) \neq h(G_2)$ implies that $G_1$ is not isomorphic to $G_2$. If the hash is fast to generate and creates small enough bags of balls, then it greatly speeds up the enumeration task. We used the tuple of the diameter and the sorted eigenvector centrality of the graph adjacency matrix to hash our balls. We manage to enumerate the 930449 balls of $\mathcal{B}_{3}$ and the regular ones in $\mathcal{B}_{2}$ in less than a day.

In order to compute each value $\langle O_X\rangle_B$, we relied on a full statevector representation using on a standard integrator (boost's ODEINT solver used inside a proprietary code). The simulation were performed on a Eviden Qaptiva appliance using a mix of CPUs and GPUs. The simulations of the 930449 balls took approximately one month. The github code offers an alternative Qutip implementation of the same numerical simulations, for reproducibility \footnote{The Open Source code for reproducibility of this work is made available to readers on GitHub \href{https://github.com/Arts-Braido/LR-bound-for-approximation}{https://github.com/Arts-Braido/LR-bound-for-approximation}. }.

\bibliographystyle{unsrt}
\bibliography{biblio}

\begin{thebibliography}{10}

\bibitem{Apolloni_1989}
B.~Apolloni, C.~Carvalho, and D.~{de Falco}.
\newblock Quantum stochastic optimization.
\newblock {\em Stochastic Processes and their Applications}, 33(2):233--244,
  1989.

\bibitem{Kadowaki_1998}
Tadashi Kadowaki and Hidetoshi Nishimori.
\newblock Quantum annealing in the transverse ising model.
\newblock {\em Phys. Rev. E}, 58:5355--5363, 1998.

\bibitem{PhysRevA.108.042403}
Shunta Arai, Hiroki Oshiyama, and Hidetoshi Nishimori.
\newblock Effectiveness of quantum annealing for continuous-variable
  optimization.
\newblock {\em Phys. Rev. A}, 108:042403, 2023.

\bibitem{Yarkoni_2022}
Sheir Yarkoni, Elena Raponi, Thomas B{\"a}ck, and Sebastian Schmitt.
\newblock Quantum annealing for industry applications: introduction and review.
\newblock {\em Reports on Progress in Physics}, 85, 2021.

\bibitem{farhi2000quantum}
Edward Farhi, Jeffrey Goldstone, Sam Gutmann, and Michael Sipser.
\newblock Quantum computation by adiabatic evolution.
\newblock Preprint at \url{https://arxiv.org/abs/quant-ph/0001106}, 2000.

\bibitem{albash2018adiabatic}
Tameem Albash and Daniel~A. Lidar.
\newblock Adiabatic quantum computation.
\newblock {\em Rev. Mod. Phys.}, 90:015002, January 2018.

\bibitem{farhi2014quantum}
Edward Farhi, Jeffrey Goldstone, and Sam Gutmann.
\newblock A quantum approximate optimization algorithm.
\newblock Preprint at "\url{https://arxiv.org/abs/1411.4028}, 2014.

\bibitem{Farhi2022quantumapproximate}
Edward Farhi, Jeffrey Goldstone, Sam Gutmann, and Leo Zhou.
\newblock The {Q}uantum {A}pproximate {O}ptimization {A}lgorithm and the
  {S}herrington-{K}irkpatrick {M}odel at {I}nfinite {S}ize.
\newblock {\em {Quantum}}, 6:759, July 2022.

\bibitem{Herrman2021MultiangleQA}
Rebekah Herrman, Phillip~C Lotshaw, James Ostrowski, Travis~S Humble, and
  George Siopsis.
\newblock Multi-angle quantum approximate optimization algorithm.
\newblock {\em Scientific Reports}, 12(1):6781, 2022.

\bibitem{shaydulin2019}
Ruslan Shaydulin and Yuri Alexeev.
\newblock Evaluating quantum approximate optimization algorithm: A case study.
\newblock In {\em 2019 tenth international green and sustainable computing
  conference (IGSC)}, pages 1--6. IEEE, 2019.

\bibitem{lykov2023sampling}
Danylo Lykov, Jonathan Wurtz, Cody Poole, Mark Saffman, Tom Noel, and Yuri
  Alexeev.
\newblock Sampling frequency thresholds for the quantum advantage of the
  quantum approximate optimization algorithm.
\newblock {\em npj Quantum Information}, 9(1):73, 2023.

\bibitem{Pelofske_2023}
Elijah Pelofske, Andreas B{\"a}rtschi, and Stephan Eidenbenz.
\newblock Quantum annealing vs. qaoa: 127 qubit higher-order ising problems on
  nisq computers.
\newblock In {\em International Conference on High Performance Computing},
  pages 240--258. Springer, 2023.

\bibitem{blekos2023review}
Kostas Blekos, Dean Brand, Andrea Ceschini, Chiao-Hui Chou, Rui-Hao Li, Komal
  Pandya, and Alessandro Summer.
\newblock A review on quantum approximate optimization algorithm and its
  variants.
\newblock Preprint at \url{https://arxiv.org/abs/2306.09198}, 2023.

\bibitem{benchasattabuse2023lower}
Naphan Benchasattabuse, Andreas Bärtschi, Luis~Pedro García-Pintos, John
  Golden, Nathan Lemons, and Stephan Eidenbenz.
\newblock Lower bounds on number of qaoa rounds required for guaranteed
  approximation ratios.
\newblock Preprint at \url{https://arxiv.org/abs/2308.15442}, 2023.

\bibitem{braida2022constant}
Arthur Braida, Simon Martiel, and Ioan Todinca.
\newblock On constant-time quantum annealing and guaranteed approximations for
  graph optimization problems.
\newblock {\em Quantum Science and Technology}, 7(4):045030, September 2022.

\bibitem{banks2023rapid}
Robert~J. Banks, Dan~E. Browne, and P.A. Warburton.
\newblock Rapid quantum approaches for combinatorial optimisation inspired by
  optimal state-transfer.
\newblock {\em {Quantum}}, 8:1253, February 2024.

\bibitem{hastings2019classical}
Matthew~B Hastings.
\newblock Classical and quantum bounded depth approximation algorithms.
\newblock Preprint at \url{https://arxiv.org/abs/1905.07047}, 2019.

\bibitem{lieb1972finite}
Elliott~H Lieb and Derek~W Robinson.
\newblock The finite group velocity of quantum spin systems.
\newblock {\em Communications in mathematical physics}, 28(3):251--257, 1972.

\bibitem{chen2023speed}
Chi-Fang~(Anthony) Chen, Andrew Lucas, and Chao Yin.
\newblock Speed limits and locality in many-body quantum dynamics.
\newblock {\em Reports on Progress in Physics}, 86(11):116001, September 2023.

\bibitem{Halperin2002MAXCI}
Eran Halperin, Dror Livnat, and Uri Zwick.
\newblock Max cut in cubic graphs.
\newblock {\em Journal of Algorithms}, 53(2):169--185, 2004.

\bibitem{Wang_2020}
Zhiyuan Wang and Kaden~R.A. Hazzard.
\newblock Tightening the lieb-robinson bound in locally interacting systems.
\newblock {\em PRX Quantum}, 1:010303, September 2020.

\bibitem{Haah_2021}
Jeongwan Haah, Matthew~B. Hastings, Robin Kothari, and Guang~Hao Low.
\newblock Quantum algorithm for simulating real time evolution of lattice
  hamiltonians.
\newblock {\em SIAM Journal on Computing}, 52(6):FOCS18--250--FOCS18--284,
  2023.

\bibitem{Barahona1988}
Francisco Barahona, Martin Gr\"{o}tschel, Michael J\"{u}nger, and Gerhard
  Reinelt.
\newblock An application of combinatorial optimization to statistical physics
  and circuit layout design.
\newblock {\em Operations Research}, 36(3):493--513, 1988.

\bibitem{Bravyi_2006}
S.~Bravyi, M.~B. Hastings, and F.~Verstraete.
\newblock Lieb-robinson bounds and the generation of correlations and
  topological quantum order.
\newblock {\em Phys. Rev. Lett.}, 97:050401, July 2006.

\bibitem{MaxCutp2}
Jonathan Wurtz and Peter Love.
\newblock Maxcut quantum approximate optimization algorithm performance
  guarantees for p greater than 1.
\newblock {\em Phys. Rev. A}, 103:042612, 2021.

\end{thebibliography}

\end{document}